\documentstyle[aps,preprint,tighten]{revtex}
\input myepsf.sty
\input myepsfig.sty
\def\ati{\alpha_{\rm TI}}

\def\equ(#1){Eq.~(\ref{#1})}
\def\pinax(#1,#2,#3,#4){\left(\matrix{#1 & #2\cr 
                                      #3 & #4\cr}\right)}
\def\tria(#1,#2,#3,#4,#5,#6,#7,#8,#9)
      {\left(\matrix{#1 & #2 & #3 \cr 
      #4 & #5 & #6 \cr 
      #7 & #8 & #9 \cr}\right)}
\def\bra(#1,#2){\left(\matrix{#1 \cr #2 \cr}\right)}
\def\ket(#1,#2){\left(\matrix{#1 & #2 \cr}\right)}
\def\equ(#1){Eq.~(\ref{#1})}

\def\Re{\mbox{Re}}

\def\nnn{\nonumber}

\def\bea{\begin{eqnarray}}
\def\eea{\end{eqnarray}}
\def\be{\begin{equation}}
\def\ee{\end{equation}}

\def\myr(#1){Eq.\ (\ref{#1})}
\def\fact(#1,#2){\scriptstyle{#1\choose #2}}

\def\a{\alpha}

\begin{document}  


\preprint{\vbox{accepted for publication in  Physical Review\ D
 \hfill NIKHEF-96-005 \\
                \null\hfill\ ADP-96-37/T234 \\ }}

\title {\bf Scaling and confinement aspects of 
         tadpole improved SU(2) lattice gauge theory
         {\bf and its abelian projection}}

\author{ Grigorios I. Poulis$^{a,b}$
       } 
\address { 
          $^a$  Department of Physics and Mathematical 
                Physics\\
                and Special Research Center for the Subatomic 
                Structure of Matter\\
                University of Adelaide, 
                SA 5005, Australia~\thanks{present address}
             \thanks{e-mail: gpoulis@physics.adelaide.edu.au}\\
            and\\  
         $^b$ National Institute for Nuclear Physics 
              and High Energy Physics (NIKHEF)\\
              P.O. Box 41882, 1009 DB Amsterdam, The Netherlands\\[2ex]
         }
\date{\today}

\maketitle       
   
\begin{abstract}{\em }

Using a tadpole improved SU(2) gluodynamics action, the
nonabelian potential and the abelian potential
after the abelian projection are computed. Rotational
invariance is found restored at coarse lattices both
in the nonabelian theory and in the effective abelian 
theory resulting from maximal abelian projection. Asymptotic 
scaling is tested for the SU(2) string tension. Deviation 
of the order of $6\%$ is found, for lattice spacings
between 0.27 and 0.06 fm. Evidence for asymptotic scaling 
and scaling of the monopole density in maximal abelian 
projection is also seen, but not at coarse lattices. The 
scaling behavior is compared with analyses of Wilson action 
results, using bare and renormalized coupling schemes.   
Using extended monopoles, evidence is found that the gauge
dependence of the abelian projection reflects short distance
fluctuations, and may thus disappear at large scales.

\end{abstract}

\draft

\section{Introduction}

Considerable progress has been achieved recently
in lattice QCD as a result of combining
two relatively old ideas: (I) improving the continuum limit 
behavior of the lattice action by adding terms 
that cancel the leading finite lattice spacing corrections 
(``Symanzik improvement''~\cite{Sym}) 
and (II) identifying a renormalized coupling which connects
lattice perturbation theory and MC simulations~\cite{Parisi,LM}.
The key observation is that the disagreement
between MC results and lattice perturbation theory
can, to a large extent, be attributed to scale-independent
(tadpole) renormalizations of the bare coupling.
By converting lattice perturbation expansions in the bare coupling
to ones using a renormalized coupling that effectively takes 
these tadpole corrections into account, lattice perturbation 
theory with the Wilson plaquette action can be reconciled 
with results from simulations in the scaling 
region~\cite{LM}. Moreover, tests of asymptotic scaling of physical 
quantities are much more successful when the perturbative
beta function is computed using such a renormalized 
coupling~\cite{KarP,F,Wup,LM}.
At the same time, the above observation suggests a mean-field
type modification of the relation between lattice links and 
continuum gauge fields~\cite{Parisi,LM}, which implies that
the leading order coefficients of the correction terms
to effect Symanzik improvement have been significantly underestimated, 
the more so, the coarser the lattice\cite{ADL}. 
Thus, besides improving lattice perturbation theory,
the effectiveness of lattice gluodynamics simulations 
{\it per se} can be dramatically enhanced by working 
at coarse lattices and using a tadpole improved version 
(henceforth referred to as ``Cornell'' action~\cite{ADL}) 
of the continuum limit improved action of L\"uscher and 
Weisz~\cite{LW}. The effectiveness of the method can be 
demonstrated, e.g., by computing the off-axis interquark 
potential at small $(6^4)$, coarse lattices (lattice spacing $a\simeq 
0.4$ fm) in  SU(3) pure gauge theory. The violation 
of rotational symmetry inherent in the Wilson 
action, manifest in the ${\cal O}(\le 40\%)$ deviation 
of the off-axis potential from a linear-plus-Coulomb 
fit to the on-axis values, is reduced to
${\cal O}(\le 15\%$) using the continuum limit improved
L\"uscher-Weisz action, and is 
essentially eliminated (of order a few \%) by 
using the continuum limit plus tadpole improved action~\cite{ADL}.

In the present work we apply these ideas in SU(2) 
pure gauge theory, focusing on confinement related
aspects in the framework of abelian projection (AP)
picture~\cite{Hooft,KronNP}. After partial gauge fixing
the original SU(N) gauge symmetry is reduced to 
the U(1)$^{N-1}$ largest abelian (Cartan) subgroup. Under 
this residual group, diagonal gluon components transform 
as abelian ``photons'', off-diagonal gluons and quarks 
as doubly- and singly-charged matter fields, respectively.
The effective abelian theory (APQCD) that results from the
integration over the off-diagonal gluons contains 
a complicated assortment of abelian Wilson loop operators of 
various sizes and charges, which describe the dynamics
of the abelian photons~\cite{KenU,Misha}, and, furthermore,
mass terms for monopoles of different sizes and shapes
~\cite{Suz-monaction}. These monopoles are 
identified as singularities in the gauge-fixing condition. 
The conjecture is then that condensation of these 
abelian monopoles leads to confinement, in the spirit of
the dual superconductor confinement mechanism
in compact QED~\cite{cqed,dGT}.

One question that we wish to address in this work is whether 
the improvement of the non-abelian 
action leads as well to improved continuum limit 
behavior of the effective abelian theory resulting 
from the projection. We do this by computing 
the on- and off-axis potential from abelian Wilson loops in APQCD and 
comparing the violation of rotational symmetry between the APQCD 
resulting from using the Wilson action (APQCD-W) and APQCD resulting 
from using the tadpole improved action (APQCD-I). 
We find that the off-axis abelian potential shows restoration of 
rotational invariance as well, allowing (at least in principle)
to study the abelian projection in small, coarse lattices.

The abelian projection is gauge-dependent. 
Early studies~\cite{KronNP} using 
local projections, e.g. diagonalizing
an adjoint operator, did not seem to support
't Hooft's conjecture. One evidently 
successful projection is the maximal 
abelian (MA)~\cite{KronPL,rev_MA}, corresponding in 
the continuum to $D_\mu^n A_\mu^{ch} \equiv \partial^\mu 
A^{ch}_\mu - ig_0[A^n_\mu,A^{ch}_\mu]$ = 0, where gauge field $A$ is
decomposed in neutral($n$) and charged($ch$) components, 
$ A_\mu$ =  $A_\mu^n$ +  $A_\mu^{ch}$.
One nice property of MA projection is that the abelian 
monopole density is consistent with asymptotic 
scaling behavior in both three~\cite{Born3d,PTW} and 
four~\cite{Born4d,Debbio4d,Suz_sc} dimensional
SU(2), which suggests it may be a physical quantity.  However, 
the evidence in four dimensions is not indisputable, essentially
because of the lack of a dimensionful parameter in pure gauge
QCD$_4$. Scaling behavior has not been observed in other projections.

Thus, the second issue that we wish to address is whether renormalized
perturbation theory and tadpole improvement allow to make more
conclusive tests of asymptotic scaling of the monopole density.
We first use renormalized perturbation theory to reanalyze
the results which have been obtained with the Wilson action.
In agreement with SU(3) results~\cite{LM} and also with
earlier SU(2) analyses~\cite{F,Wup} using the ``energy'' scheme
coupling, the asymptotic scaling behavior of the string tension 
shows remarkable improvement when using the ``potential'' scheme 
renormalized coupling~\cite{LM}. In the case of the monopole 
density, however, although we do find that the degree of 
scaling violation is reduced when the renormalized coupling
is used, asymptotic scaling is clearly violated for lattice spacings 
$a > 0.1$ fm, independently of the coupling (bare
or renormalized) used. We also find evidence for
(non asymptotic) scaling of the monopole density against 
the string tension, which clearly breaks down 
for  $a > 0.1$ fm, as well.

We then study the string tension of the improved theory
(QCD-I) and the monopole density in its abelian projection
(APQCD-I). We find deviations from scaling to be typically 
of the same order of magnitude as with the Wilson action.
Monopole properties are also similar with those in the projected
Wilson theory. In particular, we see a small scaling window
for the monopole density in MA gauge setting in 
at $\beta=3.3$ where the lattice spacing is quite
small $a\simeq 0.1 $ fm, and that only for relatively
large lattices ($L\ge 12$). We also find 
good evidence for (non-asymptotic) scaling of the 
monopole density, although not at coarse lattices. 
Thus, the improvement programme reveals that 
the monopole definition is plagued by certain
lattice artifacts that do not allow to work 
at coarse lattices.

A final objective of this work is to shed some light on the issue of
gauge dependence of the abelian projection, along the spirit of 
previous work~\cite{PTW} in three dimensions.  There, it was shown
that the difference between MA and local projections can be 
attributed to  highly-correlated short
distance fluctuations. Confinement, being a large-scale phenomenon,
should be oblivious to such fluctuations. It was shown that by 
considering monopoles defined non-locally on the lattice 
(extended monopoles) such fluctuations are averaged over, 
resulting in a converging behavior between MA and local gauges.
In the last part of this work, we apply similar considerations
in the four dimensional theory, using the tadpole improved action. 
We find a similar converging behavior of the density of extended 
monopoles between MA and F12 gauge as a function of the lattice 
size $m$ of the extended cube used to define the extended monopoles.
The weakening of gauge-dependence at large physical scales is
also demonstrated by showing that the density of extended 
monopoles in  physical units forms a universal (independent
of the lattice size of the extended monopole and the projection)
trajectory as a function of their size in physical units. These results
allow some optimism that the large-scale dynamics of the abelian
projection may after all be independent of the specific gauge
used to implement it.
 
The structure of this article is as follows: in Sec. II the action 
and the observables considered in this work are described. Results are
presented in Sec. III and our conclusions appear in Sec. IV.

\section{Method}
The action used in this work is a tadpole improved version 
of the tree-level continuum limit improved SU(2) action of
L\"uscher and Weisz. We begin by briefly summarizing the 
action improvement program. The Wilson action for SU(N) Yang Mills 
reads $S[U] = \beta\sum_{pl} S_{pl}$, where   
\begin{equation}\label{app1}
        S_{pl}\ \equiv \ {1\over N}{\rm Re{\;}Tr}(1-U_{pl})
        \ = \ {g_0^2\over 2N} a^4
        {\rm Tr}(F_{12}^2)\ +\ {\cal O}(a^6) \ .
\end{equation}
Here $g_0^2=4\pi\alpha_0$ is the bare lattice coupling constant, $a$
the lattice spacing, and we have taken for simplicity the plaquette 
$U_{pl}$ to be in the (1,2) plane. The continuum action is recovered
by identifying $\beta=2N/g_0^2$.
To improve the continuum limit behavior of the theory
(``Symanzik-improvement''~\cite{Sym})
one adds operators that correct for the  ${\cal O}(a^6)$ 
terms. Among other possible choices~\cite{Leiden} one can
use $1\times 2$ rectangular Wilson
loop (labeled ``$rt$'') and $1\times 1\times 1$ parallelogram Wilson 
loop (labeled ``$pg$'') terms~\cite{LW}
\begin{equation}\label{app2}
	S[U] \ = \ \beta \ c_{pl} \sum_{pl} S_{pl} 
	     \ + \ \beta \ c_{rt} \sum_{rt} S_{rt} 
	     \ + \ \beta \ c_{pg} \sum_{pg} S_{pg} 
                  \ , 
\end{equation} 
where the sums extend over all lattice points and relevant
orientations of the operators. To first nontrivial
order in perturbation theory, $c_i = c_i^0 + 4\pi\a{}_0\Delta_i$, 
the action in \myr(app2) reproduces the continuum action
up to and including ${\cal O}(a^6)$ terms, provided
$c_{pl}^0=5/3$, $c_{rt}^0=-1/12$, $c_{pg}^0=0$
(at tree level the coefficients are independent of 
the specific gauge group and the 
space-time dimensionality~\cite{WW}). One loop corrections 
$\Delta_i$ have been computed by L\"uscher and Weisz
for both SU(2) and SU(3) (Table 1 in Ref~\cite{LW}).

Following the convention of Ref.~\cite{ADL} we set
$c_i\beta \rightarrow \beta_i$ and redefine
$\beta\equiv \beta_{pl}$, which makes
the tree-level coefficient of the $1\times 2$ term $-1/20$~\cite{redef}. 
Given $\beta$ (which implicitly determines the strong coupling)
the other two couplings are perturbatively renormalized
\begin{eqnarray}\label{app3}
    \beta_{rt} &=& -{\beta\over 20}\biggl[1-\Bigl({3\over 5}\Delta_{pl}
                                                  + 12\Delta_{rt}
                                            \Bigr)4\pi\alpha_0
                                   \biggl] \nonumber \\
    \beta_{pg} &=& {3\over 5}\beta\Delta_{pg}4\pi\alpha_0 \ .
\end{eqnarray}

As described in Ref.~\cite{ADL}, the continuum limit behavior 
of the L\"uscher and Weisz action can be further 
improved by making the lattice links more 
``continuum like''. At the mean field level this entails setting
 $U_\mu\rightarrow u_0^{-1}U_\mu$, where one possible
choice for the mean field factor $u_0$ is using the expectation
value of the average plaquette
\begin{equation}\label{u0def}
u_0 = \langle{1\over N}{\rm Re{\;}Tr}U_{pl} \rangle^{1/4} \ .
\end{equation}
The average plaquette with Wilson action
has been calculated in lattice perturbation
theory to ${\cal O}(\a^2)$~\cite{G81} and recently to
 ${\cal O}(\a^3)$~\cite{Xarhs}.
It has also been calculated using the L\"uscher and Weisz action, 
Eq.\ (\ref{app2}), by Weisz and Wohlert~\cite{WW}. However, in the 
latter case the result in numerically known to first order only:
\begin{eqnarray}
    -\log\langle{1\over N}\rm{Re{\;}Tr}U_{pl}\rangle 
                 &=&   \xi_N\;\a{}_s \ , \label{app5a} \\
                 \xi_N  =  0.366262\; \pi {N^2\!-\! 1\over N}
                   &=&     \left\{
                     \begin{array}{ll}
		         1.72597,\quad  & \mbox{for N=2}\smallskip
                        \\
                         3.06839,\quad & \mbox{for N=3}
                     \end{array}\right.
                      \nnn \ . 
\end{eqnarray}
The L\"uscher and Weisz action can now be tadpole improved by 
explicitly pulling a $u_0^{-1}$ factor out of each link
and replacing $\alpha_0$ with a nonperturbatively renormalized 
coupling $\alpha_s$ {\em defined} through 
\myr(app5a). Since $U_{pl}$ involves 4 links and $U_{rt}$, $U_{pg}$ 
6 links, one further redefines $\beta \rightarrow \beta u_0^{-4}$,
and to recover the correct continuum limit, the relative weight of 
the correction terms is readjusted by $u_0^2=[1-\xi_N\a{}_s/2]$,
using Eq.\ (\ref{app5a}):
\begin{eqnarray}\label{app6}
    \beta_{rt} &=& -{\beta\over 20 u_0^2}
                   \left[1-\left({3\over 5}\Delta_{pl}
                                  + 12\Delta_{rt} 
                                  +  {\xi_N \over 8\pi}\right) 
                                            4\pi\alpha_s
                                   \right] \nonumber \\
    \beta_{pg} &=& {3\over 5 u_0^2 } \beta\Delta_{pg} 4\pi\alpha_s 
                          \ .
                          \end{eqnarray}
Using Table 1 of Ref.\ \cite{LW} we recover for SU(3) the improved 
action of Ref.~\cite{ADL}, while for SU(2) we find
\begin{equation}\label{app7}
    S =  \beta\sum_{pl}S_{pl} 
 -{\beta\over 20 u_0^2}
                   \left[1+0.2227\alpha_s \right]
\sum_{rt}S_{rt}
 -0.02224 {\beta\over  u_0^2} \alpha_s 
\sum_{pg}S_{pg}
                          \ .
\end{equation} 
The success of tadpole improvement can be seen in the 
value of the one-loop correction to the coefficient of
the rectangular term, which for SU(2) becomes 
$(1.08573-0.86298)$ $\rightarrow$ $0.2227$, 
a quarter of the original value. This is similar to what happens
in SU(3), where it was shown that the difference 
between results obtained using the one-loop corrected 
and tree-level tadpole improved actions is insignificant~\cite{ADL}.
The results reported here are in fact obtained from
simulations using  tree-level 
improvement only, i.e., 
\begin{equation}\label{TI_tree}
    S =  \beta\sum_{pl}S_{pl} 
 -{\beta\over 20 u_0^2}
                  \sum_{rt}S_{rt}
                          \ .
\ee
For recovering the correct continuum limit  it is important to
realize that, in the convention of Ref.~\cite{ADL} that we follow here,
the relationship between the bare coupling $\a_0$
and the simulation parameter $\beta$ in \equ(TI_tree) 
must be modified to take into account the absorption of
$c^0_{pl}$ in $\beta$~\cite{redef}
\be\label{proper}
\a_0 = 
{5\over 3}\ {N\over 2 \pi\beta} \ ,  
\end{equation}
while an additional factor $(1+1.01938\ \a_0)$ is needed in
the case of  the  one-loop corrected action of 
\equ(app7) to account for the one-loop corrected 
$c_{pl}$~\cite{Leiden}. Arguably, the RHS of 
\equ(proper) should be furthermore divided by $u_0^4$.
Following Ref.~\cite{Leiden}, this is not done here;
instead, we will denote the coupling resulting from 
such a division as a ``tadpole-improved''
or ``boosted'' coupling, $\a_{\rm TI}=\a_0 u_0^{-4}$,
as in the case of the Wilson action.

To simulate \equ(TI_tree) we thermalize using the heatbath 
algorithm, beginning with a few steps where $u_0$ is kept 
fixed to $1$. After obtaining a first estimate of the
average plaquette (and therefore for $u_0$, cf. \equ({u0def})),
we thermalize a few thousand times with $u_0$ computed every $2-4$
updates and then fed into the action, until $u_0$ stabilizes within a
few parts in $10^{-5}$. For extracting the on- and off-axis SU(2)
potential, we compute temporal $T\times{\cal C}$ Wilson loops. 
Here ${\cal C}$ are spatial paths of three types: 
(a) straight-line spatial paths of up to 6 links, from 
which we extract the $R=1,2\dots 6$ on-axis 
potential, (b) planar spatial paths ${\cal C}=1\times 2$, $1\times 3$, 
from which the off-axis  potential at $R=\sqrt{5},\sqrt{10}$ is extracted, 
and (c) cubic spatial paths ${\cal C}=1\times 1\times 1$ from which the
$R=\sqrt{3}$ potential is obtained. In the case of non-straight line
spatial paths we sum over the possible combinations allowed given the 
edges of the spatial path so as to obtain the lowest energy ($J=0$) state. 
Retaining only those paths minimally deviating from the diagonal,
there are 2 such paths for (b) and 6 for (c).  
Measurements are separated by 20-100 heatbath updates.

We then perform the abelian projection to our nonabelian configurations
(we henceforth restrict our attention to SU(2)).
A lattice implementation of abelian projection was
formulated in Refs. \cite{KronNP,KronPL}, in which
several gauge-fixing conditions were also developed 
(following 't~Hooft \cite{Hooft}). Local (generally 
nonrenormalizable) projections can be defined by the 
diagonalization of an adjoint operator~\cite{KronNP}.
Examples are diagonalization of a plaquette or a 
Polyakov line \cite{KronNP}. The maximal abelian 
(MA) projection~\cite{KronPL}
corresponds in the continuum limit to the 
renormalizable differential gauge
$D^3_\mu  A_{\mu}^\pm = 0$,
where $A^\pm_\mu \equiv (A^1_\mu \pm i A^2_\mu) / \sqrt2$.
Parametrizing the SU(2) links in the form~\cite{Suz_ext,Misha} 
\begin{equation}\label{ale3}
    U_{x,\hat\mu} = \pinax(  \cos\phi_{x,\hat\mu} \; e^{i\theta_{x,\hat\mu} }
                       , \sin\phi_{x,\hat\mu} \; e^{i\chi_{x,\hat\mu} },
      -\sin\phi_{x,\hat\mu}  \;e^{-i\chi_{x,\hat\mu} }, \cos\phi_{x,\hat\mu} \; 
      e^{-i\theta_{x,\hat\mu} } ) \ ,
\end{equation}
with $\phi\in [0,\pi/2]$ and $\chi$, $\theta$
$\in [-\pi,\pi]$, MA projection amounts to 
making the transformed links $U'_{x,\hat\mu}$ as 
close to the identity as possible
\begin{equation}\label{om}
	{\rm max}  \sum_{x,\mu}\cos(2\phi'_{x,\hat\mu}) \ .
\ee
Under diagonal SU(2) transformations 
the phases $\theta$ and $\gamma\equiv\chi+\theta$ transform 
like abelian gauge field and charge-two matter field
(in the continuum), respectively, while $\phi$ remains 
invariant~\cite{Misha}. \equ(om) is enforced iteratively;
to speed convergence (typically by a factor of 3) 
we use the overrelaxation algorithm of 
Ref.\ \cite{Mand} with parameter $\omega=1.7$.
The iteration is  repeated until the gauge transformation 
$G_x$ becomes sufficiently close to the identity
at all sites 
\begin{equation}
   {\rm max} \{ 1- {\case12} {\rm Tr}\,G_x \} \le \delta \ll 1 ,
\label{crit}   
\end{equation}
with $\delta = {\cal O}(10^{-7})$ used as a stopping criterion.

The abelian potential
after the projection is obtained from singly charged 
abelian Wilson loops constructed from the phases $\theta$
\begin{equation}\label{e10}
        W_{T\times {\cal C}}^{abel}= \Re \left\{ 
       \prod_{i\in {T\times {\cal C}} } 
              e^{i\theta_i} \right\}
                = \cos\Bigl( \sum_{i\in {T\times {\cal C}}} \theta_i\Bigr) 
               \equiv \cos\theta_{T\times {\cal C}} \ ,
\end{equation}
with the same choice of spatial paths ${\cal C}$ 
as for the SU(2) Wilson loops discussed above. 

Monopoles are identified in the abelian configurations 
using the algorithm of DeGrand and Toussaint~\cite{dGT}. Firstly, 
reduced abelian plaquette angles $\widetilde\theta_{pl}$ are defined
\begin{equation}
   \widetilde\theta_{pl} \equiv
   \theta_{pl} - 2 \pi N_{pl} , 
   \quad \widetilde\theta_{pl} \in (-\pi, +\pi] ,
\label{plaqangle}
\end{equation}
where $N_{pl}$ is identified with the number of Dirac strings 
passing through the plaquette 
($N_{pl} \in \{0,\pm 1,\pm 2\}$). The net flux of monopole current 
out of the ``elementary'' (that is, of size $1^3$) cube $C(n,\mu)$, 
 labeled by the dual lattice link $(n,\mu)$, is equal to 
the sum of Dirac strings $N_{pl}$ passing through the 
oriented $1\times 1$ plaquettes on the surfaces of 
the cube~\cite{KronNP}
\begin{equation}
   N_{m=1}(n,\mu) = - \sum_{pl} N_{pl} \ .
   \label{Kdef}
\end{equation}
We also consider type-II~\cite{IvanExtend} extended monopoles
$N_m$ constructed as the number of elementary ($m=1$) monopoles 
minus antimonopoles in a spatial cube of size $m^3$. For the
lattice density of monopoles we have adopted the definition of
Ref.~\cite{Debbio4d}
\begin{equation}
   \rho^{[m]}_{lat} =  {1\over L_t L_s^3 } \sum_{n} \vert N_{m} 
(n,4)\vert \ ,
\label{defG}
\end{equation}
i.e., the three-dimensional 
density of the time ($\mu=4$) components of the monopole
currents, averaged over all time slices $L_t$.

\section{results}

\subsection{The abelian potential}

We first discuss the continuum limit behavior of
the effective abelian theory after the projection. 
In Fig.\ \ref{W6b17}
we show the on- and off-axis QCD potential with Wilson
action (QCD-W) and the abelian potential resulting from its maximal
abelian projection (MAQCD-W). The results have been obtained
from 3100 configurations on a $6^4$ lattice
at $\beta=1.7$, with measurements separated by 40-100 updates.
 A problem common to both Wilson and improved
actions when working at coarse ($a\simeq 0.4$ fm)
lattices is the difficulty in establishing plateaus in the 
time direction for the correlators, since after $T=2$
time slices (corresponding to 0.8 fm) the S/N ratio 
has dropped dramatically.
In this work we follow Ref.~\cite{Morn1} and evaluate the potential at
$T=2$~\cite{anis}. The on-axis potential is fitted to an ansatz
$V(r)=\sigma r - \pi/ 12 r + c$ (dotted line). 
To set the scale we adopt the familiar practice~\cite{MT,Wup}
of using the physical string tension, $a\sqrt{\sigma_{N_c=2, N_f=0}}$ 
= $\sqrt{\sigma_{\rm phys}}$ $\simeq 0.44$ GeV, which suggests
a lattice spacing $a\simeq 0.39$ fm, corresponding to the
fit value $\sigma=0.75$. The large deviation 
of the off-axis points from the on-axis fit shows clearly the 
violation of rotational symmetry plaguing both the  
Wilson theory and its abelian projection at coarse lattices.
Results for the tadpole improved action (QCD-I) and its maximal 
abelian projection (MAQCD-I) are shown in Fig.\ \ref{I6b24}. They 
come from 3200 measurements on a $6^4$ lattice at $\beta=2.4$. 
A similar fit to the potential at $T$=2 suggests $a\simeq 0.39$ fm. 
A more careful estimation of the string tension using 
APE smearing~\cite{fuzz} shows that the lattice spacing has been 
overestimated by $\simeq 5\%$, and is rather close to 0.37 fm, 
which is, nevertheless, sufficiently coarse for our purposes.
The QCD results are in agreement with recent 
calculations in both SU(3)~\cite{ADL} and SU(2)~\cite{How}, 
and show that the continuum limit behavior of the 
Cornell action is clearly improved, even with tree-level 
tadpole improvement only.

The new feature emerging from these results is that not only QCD, 
but the abelian projected theory as well, shows improved 
continuum limit behavior and restoration of rotational invariance,
at least in MA projection. In a generic (other than MA) projection, 
APQCD contains abelian Wilson loop operators of various 
sizes and charges. In order to Symanzik improve such an action,
the coefficients of these terms should be carefully rearranged 
and, possibly, new terms should be added. It is not obvious whether
the addition of the rectangular term in the QCD action with
the appropriate coefficient so as to effect tree-level 
Symanzik improvement will, after the partial gauge fixing,
automatically fine-tune the coefficients of the APQCD action
in such a way as to eliminate ${\cal O}(a^6)$ corrections 
in both QCD {\it and} APQCD. In the case of MAQCD-W the 
effective abelian action is dominated by an abelian plaquette
term~\cite{KenU,Misha,P95} and using the approximations in 
Ref.\ \cite{P95} one may argue that the observed restoration
of rotational invariance is due to a corresponding improved 
abelian action dominating MAQCD-I. 
In that respect, it would be interesting to test the
behavior of the off-axis abelian potential in a local projection
(e.g., field-strength gauge, F12) 
. Although we have seen some evidence
that rotational invariance is restored in  F12 gauge as well,
the F12 abelian Wilson loops are much more noisy and do 
not allow definitive conclusions to be drawn.
Using anisotropic lattices may allow to clarify this point,
as well as the issue of abelian dominance at small
and coarse lattices, which we have not addressed here.

\subsection{scaling studies}

In this section we discuss the scaling behavior
of the SU(2) string tension and the abelian monopole density. 
The string tension has dimension [length]$^{-2}$, while
(from the physical interpretation of the monopole density as
defined in \equ(defG)) the monopole density has
dimension [length]$^{-3}$. Thus, the monopole density 
in physical units reads $\rho^{[m]}$ = $\rho^{[m]}_{lat}a^{-3}$
 =  ${\rho^{[m]}_{lat}\over (\Lambda a)^3} \Lambda^3$,
while the string tension $\sigma$ = $\sigma_{lat}a^{-2}$
 =  ${\sigma_{lat}\over (\Lambda a)^2} \Lambda^2$.
For these quantities to be physical, the coefficient 
of $\Lambda^3$ in the monopole density that of $\Lambda^2$ 
in the string tension have to become constant 
(independent of $a(\beta$)) as the continuum limit is 
approached ($\beta\rightarrow\infty$). In previous scaling 
studies (with the Wilson action)  the bare coupling was 
employed to extract a scale $\Lambda$ using the 
two-loop perturbative beta function
\begin{equation}
	a(\beta)\Lambda = \left( 12\pi \over 11 N \a\right)^{51\over 121}
                     \exp\left(-{6\pi\over 11 N \a}\right) \ .
	\label{L0}
\end{equation}

Before discussing the scaling behavior with the improved action,
however, it may be  instructive to revisit the 
scaling behavior of the Wilson action results, this time using 
both bare (BPT) and renormalized (RPT) perturbation theory.


\subsubsection{ Wilson action}

In BPT, the bare coupling $\alpha_0$ = 
$(\pi\beta)^{-1}$ is used to extract the scale from \equ(L0),
while, in RPT, a renormalized coupling is employed, e.g.,
the ``energy'' coupling $\a_{\rm E} = {2N\over\pi (N^2\!-\!1)}
(1-\langle \Box\rangle)$~\cite{Parisi,KarP}, or
the ``tadpole improved'' coupling $\ati = \a{}_0 u_0^{-4}$~\cite{LM}, 
or, more effectively, the ``potential'' scheme coupling $\a{}_{\rm V}$, 
defined from the nonperturbatively computed
heavy-quark potential~\cite{LM}. In the latter case, 
instead of measuring the potential, one invokes
the lattice perturbation theory expansion of the heavy-quark 
potential~\cite{BK}
\begin{eqnarray}\label{kov}
	V(q) &=& - {N^2\!-\!1\over 2 N}{4\pi\a_0\over q^2}\left(
                          1 + \a_0\left[
         {N\over 3}{11\over 2\pi}\log\left(\pi\over a q\right)
		-{N J\over 4\pi}\right]\right) +{\cal O}(\a_0^3)\nnn\\
            &\equiv&  - {N^2\!-\!1\over 2 N}{4\pi\a_{\rm V}\over q^2} \ ,
\end{eqnarray}
where $J=-19.695$ and $-16.954$, for $N=3$ and $N=2$, respectively,
together with the analogous expansion of some other 
short-distance quantity, e.g., the average 
plaquette~\cite{G81,Xarhs}, 
\begin{eqnarray}\label{dGRos}
1-\langle{1\over N}{\rm Re{\;}Tr}U_{pl}\rangle 
&=& c1\;\a_0 + c2\;\a_0^2 +{\cal O}(\a_0^3) \nnn\\ 
c1 = {N^2\!-\!1\over 2N}\pi, &&
c2 = (N^2\!-\!1)\left({4\pi\over 2 N}\right)^2
    \left(0.0203\;N^2 - {1\over 32}\right) \ ,\nnn\\
\end{eqnarray}
in order to extract 
$\a{}_{\rm V}$ from the measured value of this quantity:
\begin{equation}
\label{Wplaq2}
	-\log\langle{1\over N}{\rm Re{\;}Tr}U_{pl}\rangle 
       ={N^2\!-\!1\over 2N}\pi\;\a_{\rm V}({q^*\over a})
               \left(
                       1 +
                \delta_N \;\a{}_{\rm V}
                \right) \ ,
\end{equation}
where
\be\label{lkj}
 \delta_N  = {11N\over 6\pi} \log({q^*\over\pi}) +
              {N J\over 4\pi} +
               {c2\over c1} + {c1\over 2}  \ .        
\ee                    
At $q^*=\pi$ one has $\delta_N$ = $-1.3386$ and $-0.8925$,
for $N=3$ and $N=2$, respectively.
According to the procedure proposed by Lepage and 
Mackenzie for fixing the scale, $q^*=3.41$~\cite{LM}. 
Another nonperturbative coupling may by obtained by solving 
\equ(Wplaq2) to first order (i.e., ignoring $\delta_N$) and
will be denoted by $\a_s$, so as to facilitate comparisons 
with the improved action [cf. \equ(app5a)]. 

Consider first the density of elementary monopoles,
whose asymptotic scaling behavior has been tested, 
using BPT, in Refs.~\cite{Born4d,Debbio4d}). The 
open circles in the inner graph of Fig.\ \ref{Wscaling}. 
correspond to the $12^4$ data in Fig.~1 of Ref.~\cite{Born4d}. 
Although the results are not incompatible with scaling, more solid 
evidence is required. The recently computed three-loop 
beta function for Wilson action~\cite{threeloop} modifies
\myr(L0) by a factor $(1+0.08324\;4\pi\a_0)$, which, 
however, does not improve the evidence for asymptotic 
scaling, since the $\beta=2.5$ and $\beta=2.6$
values come closer by less than 1\%.
Asymptotic scaling in MAQCD-W using the same data but 
renormalized perturbation theory
is tested in the main graph in Fig.\ \ref{Wscaling},
with $a\Lambda$ extracted from the $\a{}_{\rm V}$ coupling and
$q^*=3.41$. Only results with the two-loop beta function are shown,
since the (scheme dependent) three-loop correction is negligible
in renormalized coupling schemes~\cite{threeloop}. The values still do
not lie on a plateau, however, the $\beta$ dependence is significantly
reduced: between $\beta$ = 2.5 ($a\simeq$ 0.085 fm) and 2.6 ($a\simeq$
0.061 fm) the density drops by 23\% using the bare coupling, 
but by much less (9\%) using the potential coupling. 
Between $\beta$ = 2.5 and 2.7 ($a\simeq 0.045$ fm) the corresponding
numbers are 37\% and 13\% with $\a_0$ and $\a_{\rm V}$, respectively.
Reduced scaling violation is observed with other renormalized couplings 
as well, however, it is less effective compared to the potential 
scheme: between $\beta=2.5$ and $\beta=2.7$ we find 27\% scaling 
violation with $a_{\rm TI}$ and 17\% with $\a_{\rm s}$ (the 
``first-order potential'' coupling). Results from larger lattices
show  smaller scaling violation: by reanalyzing the $16^4$ 
data points in Table~2 of Ref.\ \cite{Debbio4d} we find
9.3\%, 4.5\% and 4.2\% scaling violation with $\a_0$, 
$a_{\rm TI}$ and $\a_{\rm s}$, respectively, 
between $\beta=2.5$ and $\beta=2.6$.

The information Fig.\ \ref{Wscaling} reveals 
regarding the status of the MA monopole density as a possibly 
physical quantity should be properly assessed against similar 
tests for {\it bona fide} physical quantities, such as the string tension.
Asymptotic scaling for the SU(2) string tension has been tested 
in bare PT~\cite{MT}, as well as in renormalized PT, using the 
energy scheme coupling~\cite{F,Wup}. In Fig.\ \ref{s-sc-W} we reproduce the
BPT test (inner graph) and also show results for RPT in the 
$\alpha_{\rm V}$ scheme (main graph)~\cite{beta24}. As has been 
observed in Ref.~\cite{MT}, asymptotic scaling is not satisfied 
in bare PT, although the $\beta > 2.7$ data from more recent 
calculations~\cite{F,Wup} approach a plateau.
Scaling violation is substantially reduced when renormalized PT 
is used, in agreement with Refs.~\cite{F,Wup} where the $\a_{\rm E}$
coupling was employed. Specifically, from Table~\ref{Wschemes} 
follows that between $a \simeq 0.17$ and $\simeq 0.4$ fm the 
violation of asymptotic scaling is $\simeq 27\%$ when the 
bare coupling is used, 16\%  with the tadpole 
improved coupling, 8\%  with either the energy 
coupling, $\a_{\rm E}$, or $\a_{\rm s}$, the coupling
obtained from \myr(Wplaq2) to first order,
and is reduced to  6\% with the $\a_{\rm V}$ coupling,
obtained by solving \myr(Wplaq2) to second order, 
with $q^*=3.41$. 

Moreover, combining all these calculations we find behavior 
compatible with (non asymptotic) scaling of the 
monopole density against the string tension, as is seen
in Fig.\ \ref{ratio-W} (see also~\cite{Bali}). However, this scaling clearly 
breaks down at lattice spacings larger than some critical value
somewhere  between 0.12 and 0.09 fm.

\subsubsection{Improved action}

When testing asymptotic scaling with the improved action 
of \equ(TI_tree), we again extract the scale $\Lambda$ from
\myr(L0). Given the value of $\beta$ used in the simulation, 
several choices of associated strong couplings $\a$ 
to be used with \equ(L0) are available, as with the Wilson action,
e.g., the bare coupling, \equ(proper), or the ``tadpole-improved''
coupling $\a_{\rm TI}=\a_0 u_0^{-4}$.
The situation is somehow  different with respect to the 
potential scheme coupling, $\a_{\rm V}$, since, 
as remarked in Sec. II, the expansion leading to
\myr(app5a) has been (numerically) carried out to first order
only. Thus, we do not have the analog of \equ(Wplaq2)
whose solution (to second order) would give the corresponding
$\a_{\rm V}$ coupling for the improved action. A solution 
to first order (in which case it does not matter which
scale $q^*$ is the coupling extracted at) 
is available, of course, and corresponds to the 
nonperturbative coupling $\a_{\rm s}$ of \equ(app5a).

Let us now discuss the results obtained with the improved action. 
Starting with the density of elementary $(m=1)$ monopoles
in APQCD-I, results from $8^4$, $12^4$ and  $16^4$ lattices, 
for $\beta$ values between 2.7 and 3.7, are shown  
in Fig.\ \ref{raw} (see also Table~\ref{den_run}). 
The behavior of the monopole density  is very similar
to what has been observed in earlier studies using the Wilson
action~\cite{Born4d,Debbio4d}. In particular, the raw density 
of F12 monopoles appears to be independent
of $\beta$ (and therefore of the lattice spacing), implying  
that in physical units it diverges like $a^{-3}$. On the
other hand, the MA monopoles do seem to develop an 
exponential falloff which is a necessary condition for scaling 
behavior. One also notices the considerable volume dependence
of the MA monopole density results, which should be expected
if monopoles play some role in the confinement mechanism: 
the lattice spacing is $a\simeq 0.087$ fm at $\beta=3.4$ 
(Table~\ref{fits}) and the underestimation of the density on
the $8^4$ lattices compared with the $12^4$ results 
reflects the inadequacy of the smaller lattice to
provide for the typically 1 fm confinement scale
(similarly at $\beta=3.5$, where $a\simeq 0.075$,
between the $12^4$ and $16^4$ results). 
This underestimation is also in agreement with the expectation
that it is the large monopole loops that are strongly correlated with 
confinement~\cite{Stack,HartTeper}.

Asymptotic scaling for the monopole density in MAQCD-I
is tested in Fig.~\ref{mono-scaling}, using the bare coupling
of \equ(proper), the boosted or tadpole-improved 
coupling $\a_{\rm TI}$, and the ``first-order potential'', 
nonperturbative coupling $\a_{\rm s}$. The results are, 
in every case, not incompatible with scaling, 
at level similar to that of the Wilson action 
results: between $\beta$ = 3.4 and 3.6
($a \simeq 0.087$ and 0.063 fm, respectively, corresponding roughly 
to the $\beta \in [2.5,2.6]$ interval for the Wilson
action discussed above) the scaling 
violation of the $16^4$ data is 8.6\%, 6.1\% and 3.4\%
using $\a_0$, $\a_{\rm TI}$ and $\a_{\rm s}$, respectively.
At even smaller lattice spacings, for fixed lattice size,
the lattice volume becomes too small and this results 
in the expected ``enveloping'' curve~\cite{Born4d}.

Turning to the string tension, this is extract from linear 
+ Coulomb chi-squared fits (without fixing the Coulomb 
coefficient to the L\"uscher value) to the time-dependent 
potential $V(R,T)$ = $\log \left(W[R,T\!-\!1]/W[R,T]\right)$, 
using jackknife errors for $V(R,T)$. 
It is known that the correction terms in the action spoil 
the hermiticity of the transfer matrix~\cite{Ltransf}. As 
a result, correlators show damped oscillatory behavior in 
$T$~\cite{anis}. Using APE smearing~\cite{fuzz}, we have 
nevertheless established plateaus in $T$ in our fits of the 
string tension (Table~\ref{fits}); results from 
the $16^4$ lattices are shown in in Table~\ref{str_run}. 
Above $\beta=3.4$ the string tension from $12^4$ lattices 
(not shown in Table~\ref{str_run}) is underestimated 
(commensurate with the deviation of the monopole density results
between $12^4$ and $16^4$ lattices discussed above),
since the lattice size is below 1 fm in this case.
This can be seen in Fig.~\ref{sigma-scaling}, where we test 
asymptotic scaling for the string tension using the improved
action, and with the same choice of couplings as for the 
monopole density. Since the three-loop correction to the beta function is not 
universal, the Wilson action computation in Ref.\ \cite{threeloop}
in not applicable in this case. The results appear compatible with
asymptotic scaling, especially when the boosted coupling, 
$\a_{\rm TI}$, is used. Specifically, from Table\ \ref{schemes} is seen
that the violation of asymptotic scaling between $\beta=2.7$ 
($a\simeq 0.26$ fm) and $\beta=3.6$ ($a\simeq 0.06$ fm) 
is $\simeq 6\%$  with $\a_{\rm TI}$, but more pronounced, 
$\simeq$ 12.5\%, using either one of 
$\a_{\rm 0}$, $\a_{\rm s}$ and $\a_{\rm E}=(1-\langle\Box
\rangle)/1.7259$ (not shown). Apparently (and unlike the monopole density
results discussed above) the nonperturbative coupling $\a_{\rm s}$
is not so successful here, compared to $\a_{\rm TI}$. However, 
this depends strongly on the value of the string tension at the 
coarser lattice considered, $\beta=2.7$: the scaling violation
between  $\beta=2.9$ ($a\simeq 0.2$ fm) and $\beta=3.6$ 
is  6\% and 4\% with $\a_{\rm TI}$ and $\a_{\rm s}$, respectively.
In general, asymptotic scaling tests using these two couplings
give rather similar results, since the ratio $\a_{\rm s}/\a_{\rm TI}$
slowly varies around 0.930 with either Wilson or improved action, as seen
from Tables~\ref{Wschemes} and~\ref{schemes}. Notice, 
though, that the ratio does not monotonically decrease with 
$\beta$, as does, e.g., for the ratio $\a_{\rm s}/\a_0$
in the case of the improved action [cf. Table\ \ref{schemes}], or, 
in the case of the Wilson action, when one instead of using 
$\a_{\rm s}$ solves \myr(Wplaq2) to second order to
obtain  $\a_{\rm V}$ (last column in Table~\ref{Wschemes}).
It is conceivable that knowledge of the second order
terms in \myr(app5a) and the improved action lattice perturbation
expansion of the potential (the analog of \equ(kov)), 
which would allow to extract the $\a_{\rm V}$ coupling with
the improved action, might lead to even smaller 
asymptotic scaling violation.

As with the Wilson action, we can now combine the results
to test (non asymptotic) scaling. Reading 
off $\sqrt{\sigma}\simeq 3.8\Lambda_{\rm TI}$ 
from Fig.~\ref{sigma-scaling}(b) and estimating
the scaling value for the monopole density 
at $\rho^{[1]}\simeq$ 54$\Lambda_{\rm TI}^3$ from 
Fig.~\ref{mono-scaling}(b)
suggests that the dimensionless combination $\rho^{[1]}\sigma^{-3/2}$
should saturate to $54\times 3.8^{-3} \simeq 1$. 
This is indeed verified in Fig.~\ref{scaling-rs}. 
By combining the $12^4$ and $16^4$ results for
the two quantities, their individual 
volume dependence cancels, to a large extent, in the
dimensionless ratio.  
Thus, the evidence for scaling from this
graph is more compelling than the evidence for asymptotic 
scaling in Fig.~\ref{mono-scaling}. 
It appears, therefore, that the elementary monopole
density in maximal abelian gauge is indeed
a physical quantity. Using $\sqrt{\sigma}=0.44$ GeV 
with $\rho^{[1]}\simeq\sigma^{3/2}$ implies a density of
approximately 11 elementary monopoles per fm$^3$, in good 
agreement with estimates using the Wilson action~\cite{Debbio4d}.
However, although scaling is observed,
its onset is at relatively small lattice spacings 
$a\simeq$ 0.1 fm, and therefore the situation 
for the monopole density is quite similar to
what we found with the Wilson action (Fig.\ \ref{ratio-W}). 

\subsubsection{Comparison of Wilson and improved action}

A comparison of the scaling results discussed above 
with the two actions suggests the following:
\begin{enumerate}
\item 
      In both cases, renormalized couplings lead to improved
      asymptotic scaling behavior compared to the bare 
      coupling\footnote{ note that this is not true for the
                           improved action if $a_{\rm TI}$ 
                           is interpreted as
                           a bare coupling (see discussion 
                            below \equ(proper)).
                        }; 
      among renormalized couplings, the nonperturbative ones 
      ($\a_{\rm s}$ or $\a_{\rm V}$, when available) are typically more
      successful than the tadpole-improved coupling $\a_{\rm TI}$.
\item 
      The string tension using the Wilson action shows more 
      pronounced scaling violation compared to using the 
      improved action, but only when analyzed with the bare coupling.
      When renormalized couplings are used, the results are 
      rather similar for the two actions, for both observables tested here
      (string tension and monopole density), i.e.,
      renormalized PT with the improved action does not 
      improve renormalized PT analyses of Wilson action results
      significantly. Thus, the results using the improved
      action lead to the same interpretation as the one 
      drawn from using a renormalized coupling with the Wilson action:
\begin{itemize}
\item 
      the string tension calculation suggests that
      the improvement program does allow confinement to be 
      studied at relatively coarse lattices.
\item
      evidently a physical quantity, the monopole density 
      in maximal abelian projection is, nevertheless, 
      sensitive to small distance ($a = 0.1$ fm) physics.
      This sensitivity persists even when a continuum
      limit improved action is used. 
\end{itemize}
In that respect, the merit of the tadpole-improvement program
(in the form of, first, renormalized perturbation theory, and
second, tadpole-improving the action itself), insofar as 
abelian monopoles are concerned, has been
to point out that the asymptotic scaling behavior
of the string tension can be extended to coarser lattices, 
while that of the  monopole density can not, 
something that --- due to the lack of asymptotic scaling
with Wilson action and with the bare coupling --- 
could not have been previously realized.
To further  clarify some of the above points, more accurate 
      determinations of the string tension with the improved     
      action using anisotropic lattices (following 
      Ref.~\cite{anis}) and, possibly, a continuum limit improved 
      version of Creutz ratios, should be performed~\cite{ImpC}.

\end{enumerate}

\subsection{Gauge dependence and extended monopoles}

It has been pointed out by several 
authors~\cite{PTW,Born4d,Debbio4d,Suz_sc,Suz_ext}
that the large number of monopoles and the associated 
scaling violation observed in local, unitary gauges (such
as F12 or Polyakov gauge) can, to some degree, be attributed
to strong short distance fluctuations. Defining monopoles
nonlocally (e.g., extended monopoles, introduced first by 
Ivanenko et al.~\cite{IvanExtend}), may help
average --at least partially--  over such fluctuations 
and therefore reduce the gauge-dependence
of the abelian projection. The main results of an analysis 
along these lines in the $d=3$ theory with Wilson action 
by Trottier, Woloshyn and this author were 
the following~\cite{PTW}:

\begin{enumerate}
	\item  monopoles in MA projection form a dilute
	        plasma in $d=3$. The monopole distribution is basically random
	        and characterized by the average minimum
	          monopole-antimonopole separation $\langle r_{\rm min}\rangle$
	        which in physical units scales. The density of
	        extended monopoles $\rho^{[m]}_{lat}$ remains roughly the
	         same for ``sizes'' $ma < \langle r_{\rm min}\rangle$
               (in physical units).

	\item  In local gauges $\rho^{[1]}_{lat}$ does not scale. 
               The distribution is significantly narrower than what 
               expected from a random distribution, indicating 
               strong short-distance correlations. However,
               unlike MA monopoles, $\rho^{[m]}_{lat}$ drops 
               rapidly with $m$, suggesting that by considering 
               extended monopoles such correlations (which are absent in MA
               projection but create ``spurious'' monopoles in, e.g., F12) 
               are being  ``washed out''. Accordingly, the
	       ratio of extended monopoles between F12 and MA projections is 
	       found to decrease as $m$ increases.
	         	
	\item For extended monopoles with fixed (physical) size, 
              $\rho^{[m]}$ (in physical units) scales for MA gauge but not for
	      local ones. However, the {\em degree of scaling violation}
	      is found to decrease substantially as the extended monopole
              size increases.
	
\end{enumerate}

To what extent do such considerations apply to the 
four-dimensional theory? The situation appears quite similar 
on a first inspection: $\rho^{[m]}_{lat}$ drops faster
in the local projections than in MA, as can
be seen by plotting the ratio $R^{[m]}\equiv$ $
\rho^{[m]}_{F12}/\rho^{[m]}_{MA}$
as a function of $m$ (Fig.~\ref{conv}).
Empirically, we find that the extended monopole 
density in local gauges (which is, in $d=4$, practically independent
of the lattice spacing for all $m$) behaves like 
 \begin{equation}
 	\rho^{[m]}_{lat}(\beta) = {C\over m^\gamma} \ ,
 	\label{f12}
 \end{equation}
with $C=0.31(1)$ and $\gamma=1.64(1)$, over a range of $\beta$
values where the lattice spacing drops by a factor of 3 
(Fig.~\ref{power}); $C$ and $\gamma$ depend very mildly on
the particular local gauge fixing, suggesting a purely
geometrical origin of $\gamma$, which, however, we have not 
identified. The MA density can also be parametrized in 
the form of \equ(f12), although the fits are not equally good.
In this case the the coefficients decrease with $\beta$,
mildly for $\gamma_{\rm MA}$ (Fig.~\ref{power})
and  exponentially for $C_{\rm MA}$ (not shown).
Since $\gamma_{\rm MA} < \gamma_{\rm F12}$,
the convergence between $\rho^{[m]}_{\rm MA}$ and
$\rho^{[m]}_{\rm F12}$ for fixed $\beta$  seen in Fig.~\ref{conv}
readily follows from this parametrization. 
The more rapid convergence for the higher $\beta$ 
values observed in Fig.~\ref{conv} is accounted for 
by $\gamma_{\rm F12}$ and  $\gamma_{\rm MA}$  being 
independent of $\beta$ and decreasing with $\beta$, respectively.

Moreover, one may show that, although, as in $d=3$, 
$\rho^{[m]}= Cm^{-\gamma}a^{-3}$ (the density in physical units)
does not scale, the degree of scaling 
violation decreases as a function of the monopole size
$ma$ in physical units. Unlike $d=3$, where $\beta$ is 
dimensionful and therefore this can be tested directly
by comparing densities with $m/\beta$ = fixed, in $d=4$
can only be implicitly deduced, from \equ(f12); 
 indeed, a measure of scaling violation,
as the continuum limit $a\rightarrow 0$ is approached, 
is given by $\partial \rho^{[m]}/ \partial a^{\epsilon}$, 
with $\epsilon < 0$. From the above parametrization
\begin{equation}
\left. {\partial\rho^{[m]}\over\partial a^{\epsilon}}\right|
                  \begin{array}{ll}
                        \\
                        _{ {ma}={\rm fixed} 
                         }
		   \end{array}     
             = C (ma)^{-\gamma} \ , \epsilon=\gamma -3 < 0 \ \ .
	\label{violation}
\end{equation}
Since $0 < \gamma < 3$, the degree of scaling violation 
for the extended monopole density in unitary gauges
will decrease as a function of the extended monopole 
size (in physical units), like in QCD$_3$~\cite{PTW}.

These results suggest that there is more contamination
from short-distance physics in the local projection,
as expected from the non-locality of the  
gauge condition in maximal abelian projection. 
In attempting to draw parallelisms with the $d=3$ case,
one should bear in mind that monopole d.o.f. are pointlike in
 $d=3$ but form closed loops in $d=4$. Thus, although a concept
of minimum separation can still be devised at $d=4$, 
it is probably not the correct way to describe
the monopole distribution. Indeed, although we find that
$\langle r_{\rm min}\rangle$ is larger in MA than in F12 
projection (by a factor ranging 
from  1.2 to 2 for the $\beta$ values we have considered),
when converted to physical units neither scales.
That may explain why $\rho^{[m]}_{lat}$ in MA projection does not
remain constant over some range in Fig\ \ref{conv}, as in $d=3$,
but instead starts to drop immediately with increasing $m$. 
The appropriate way to describe the monopole distribution is
probably by categorizing the loops according to their 
length~\cite{HartTeper}; although not discussed here,
the analog of the $d=3$ case may be to examine
how close to random the MA monopole loop length distribution 
found in Ref.~\cite{HartTeper} is. 

In order to test the behavior of monopoles at large scales
we plot in Fig.\ \ref{univ} a family of $m={\rm fixed}$ trajectories
for $\rho^{[m]}_{\rm F12}$ and $\rho^{[m]}_{\rm MA}$ as 
functions of the extended monopole size $m a$
in physical units, using the scale extracted from the $\a_{\rm TI}$
coupling. The linear behavior of the F12 data points, 
with the $m$-independent, equal to $-3$, slope and the increasing
like $\log m$ ordinates for a given abscissa, follows directly from the 
$\beta$-independence of the coefficients in \myr(f12) with $\gamma < 3$:
\begin{equation}\label{logar}
	\log\left(\rho^{[m]}_{lat}a^{-3}\right) = \log C + (3-\gamma)\log m
	-3\log\left(ma\right) \ .
\end{equation}
For each individual $m = \rm{fixed}$ trajectory,  Fig.\ \ref{univ} 
essentially tests asymptotic scaling for $\rho^{[m]}$. 
The most interesting feature is the formation of a 
universal, i.e., $m$-independent, trajectory for MA
extended monopoles of large size in {\em physical} 
units. Unlike the F12 case, the dependence of
$C_{\rm MA}$, $\gamma_{\rm MA}$ on $\beta$ (and therefore, implicitly, 
on $a$) does not allow a simple
explanation of this feature. The deviation of the 
individual $m={\rm fixed}$ curves from this universal trajectory occurs 
at $ma_c\Lambda_{\rm TI}\simeq$ 0.32 , 0.22, 0.16, 0.11, for $m=$ 6, 4, 3, 2,
respectively. Thus, the crossover points involve a common
critical lattice spacing $a_c\Lambda_{\rm TI}\simeq 0.055$, 
which, from the scaling value $\sqrt{\sigma}$ 
= 3.8$\Lambda_{\rm TI}$ = 0.44 GeV implies $a_c\simeq 0.094$ fm.
Given that these results are from $12^4$ lattices, 
 and since for fixed $m$ the continuum limit 
is to the left on the horizontal axis in Fig.\ \ref{univ}, 
it appears that this deviation from the universal 
trajectory is merely another manifestation of the
 finite physical volume effect occurring when the 
lattice size $L a$ is less than 1 fm, as described 
in~\cite{Born4d} and also observed in Sec. III.
It is quite conceivable  that towards the infinite volume limit
this universal trajectory will extend to arbitrarily small physical sizes.
We also notice that the difference between MA and F12 results 
decreases rapidly with increasing monopole size and at large physical sizes 
a projection-independent trajectory seems to form, indicating
that also in $d=4$ the abelian projection
 shows evidence of gauge independence when large scales
in physical units are probed.

\section{summary}

In this work the ideas of renormalized perturbation theory 
and tadpole improvement have been used to study confinement related 
aspects of lattice SU(2) gluodynamics.
Observables have been studied that are either directly 
related to confinement, e.g., interquark potential/string tension,
or conjectured to be in the context of a dual 
superconductor picture, e.g., monopoles and abelian 
potential after the abelian projection. 
Two types of studies have been undertaken.
Firstly, tests of the improvement program in QCD
and its abelian projection, APQCD, specifically 
(a) whether asymptotic scaling and scaling is observed,
and (b) what is the continuum limit behavior of APQCD
at small, coarse lattices. Secondly, studies of extended 
monopoles, using the tadpole improved action as a new, better tool,
with the objective of shedding some light onto the issue 
of the apparent, albeit bothersome, gauge dependence of the 
abelian projection. The results can be summarized as follows:

\begin{itemize}

\item  
      At small, coarse lattices the degree of violation of 
      rotational invariance is similar in Wilson QCD
      and its abelian projection. With the improved 
      action, though, rotational invariance is restored in both QCD 
      {\em and} the corresponding abelian theory, at 
      least in MA projection.
\item 
      Deviation from asymptotic scaling for the SU(2)
      string tension is observed at the 6\% level between 
      $a=0.06$ and 0.26 fm. This order of scaling violation
      is quite similar with renormalized coupling analyses
      of Wilson action results, although considerably improved
      in comparison to the corresponding bare coupling 
      analysis, even when
      the three-loop beta function is used. The quality of our 
      calculation does not allow statements to be made about
      scaling at the 1\% level. Using anisotropic lattices,
      will allow more accurate determination of the string 
      tension for lattice spacings $a > 0.2 $ fm. 
      However, it does not seem very likely that, even 
      with this technique, asymptotic scaling will be 
      verified at the 1\% level for coarse and small lattices. 
      The fact that continuum limit improvement is not 
      very evident in the string tension calculation, 
      is, however, not surprising, since the string 
      tension is obtained from the standard on-axis  potential.
\item
      Good evidence for scaling of the density
      of maximal abelian monopoles (using the SU(2) 
      string tension to set the scale) is found.  
      Some evidence for asymptotic scaling is seen as well, 
      although hampered by finite volume effects.
      It is, nevertheless, clear, that neither scaling nor
      asymptotic scaling hold for the MA monopole density
      at as coarse lattices as 10\% asymptotic scaling 
      for the string tension does.

\item 
      The gauge dependence of the density of abelian monopoles
      is significantly reduced when considering extended monopoles 
      of large sizes in physical units. 
\end{itemize}

The scaling studies of the monopole density
suggest that the monopole density in MA projection
is a physical quantity, and thus seem to settle a hitherto 
not entirely clarified issue. Together with the
reduced scaling violation in local projections 
at large scales, this may be considered as supporting 
evidence for the abelian projection picture of confinement.
However, despite the encouraging results from the off-axis 
abelian potential, scaling of the monopole density 
breaks down at coarse lattices. It is conceivable
that this necessitates improvement of the monopole identification
algorithm besides improving the action~\cite{RoyG}.
 
\acknowledgments
  The author would like to thank Richard Woloshyn, 
  Howard Trottier and Pierre van Baal for useful discussions 
  and suggestions, and also TRIUMF and the INT at the 
  University of Washington for their hospitality during 
  the summer of 1995. This work has been supported 
  by Human Capital and Mobility Fellowship ERBCHBICT941430
  and by the Research Council of Australia.


\begin{table}[htb]
\renewcommand{\arraystretch}{1.3}
	\caption[dummy]{Asymptotic scaling test
                  for SU(2) string tension with Wilson action.
                  Different coupling schemes are used to get $a(\beta)\Lambda$
               from \myr(L0). The bare coupling
       $\a_0  = (\pi\beta)^{-1} $, the tadpole improved coupling
       $ \a_{\rm TI}  = \a_0 \langle \Box\rangle^{-1}$, the energy coupling
       $ \a_{\rm E}  ={4\over 3\pi}(1-\langle\Box\rangle)$, and the potential coupling
      obtained from \myr(Wplaq2) with $q^*=3.41$, to first order,
       $ \a_{s}= -{4\over 3\pi}\log\langle\Box\rangle$,
      and to second order,  $\a_{\rm V}$. Data from~\cite{F,Wup} and references therein.
} 
	\protect\label{Wschemes}
\begin{tabular}{cccccccccc}
	\hline
	$\beta$  
      & $<\Box>$
      & $\sqrt{\sigma}$
      & $\sqrt{\sigma}\over\Lambda_{0}$
      & $\sqrt{\sigma}\over\Lambda_{\rm TI}$
      & $\sqrt{\sigma}\over\Lambda_{\rm E}$
      & $\sqrt{\sigma}\over\Lambda_{s}$
      & $\sqrt{\sigma}\over\Lambda_{\rm V}$
      & $ \a_{s}\over \a_{\rm TI} $
      & $ \a_{\rm V}\over \a_{\rm TI} $
      \\    
	\hline
 2.3    & 0.6024 & 0.3690(30) & 62.4( 5) & 6.59( 5) & 22.29(18)  & 2.49(2) & 1.12(1) & 0.936 & 1.199\\ \hline
 2.4    & 0.6300 & 0.2660(20) & 57.8( 4) & 6.44( 5) & 22.75(17)  & 2.55(2) & 1.16(1) & 0.931 & 1.155\\ \hline
 2.5    & 0.6522 & 0.1870(10) & 52.3( 3) & 6.03( 3) & 22.07(12)  & 2.47(1) & 1.14(1) & 0.929 & 1.127\\ \hline
 2.5115 & 0.6544 & 0.1836(13) & 52.9( 4) & 6.11( 4) & 22.41(16)  & 2.51(2) & 1.15(1) & 0.929 & 1.124\\ \hline
 2.6    & 0.6701 & 0.1360(40) & 49.0(14) & 5.76(17) & 21.52(63)  & 2.41(7) & 1.11(3) & 0.930 & 1.109\\ \hline
 2.635  & 0.6757 & 0.1208( 1) & 47.5( 1) & 5.62( 1) & 21.08( 2)  & 2.36(1) & 1.09(1) & 0.931 & 1.104\\ \hline
 2.7    & 0.6856 & 0.1015(10) & 47.1( 5) & 5.62( 6) & 21.28(21)  & 2.38(2) & 1.11(1) & 0.932 & 1.096\\ \hline
 2.74   & 0.6913 & 0.0911( 2) & 46.8( 1) & 5.61( 1) & 21.32( 5)  & 2.38(1) & 1.11(1) & 0.932 & 1.092\\ \hline
\end{tabular}
\end{table}
\begin{table}[htb]
\renewcommand{\arraystretch}{1.3}
	\caption{ Runs for monopole density determination.
 After $N_{\rm therm}$ thermalization steps,
 $N_{\rm measur}$ measurements are taken, separated by 
  $N_{\rm separ}$ updates.}
	\protect\label{den_run}
\begin{tabular}{ccccccc}
	\hline
	$\beta$  
      & lattice  
      & $N_{\rm therm}$
      & $N_{\rm measur}$ 
      & $ N_{\rm separ}$ 
      & $ < \Box >$ 
      & $\rho^{[1]}_{\rm MA}$ 
      \\    
	\hline
2.7 & $12^4$ &  $-$ & 100 & 100 & 0.5771(2) & 0.09157(27) \\
	\hline
2.9 & $12^4$ & 2500 & 500 & 60 & 0.6216(1) & 0.05295(10) \\
	\hline
3.1 & $12^4$ & 2500 & 500 & 60 & 0.6581(1) & 0.02548( 9)  \\
	\hline
3.2 & $12^4$ &  $-$    &  30 & 20  &  $-$          & 0.01641(36)  \\
    & $16^4$ & 2500 & 30 & 50  & 0.6730(1)  & 0.01706(13) \\
	\hline 
3.3 & $12^4$ & 2500 & 500 & 60 & 0.6862(1) & 0.01106( 7)  \\
    & $16^4$ & 3000 & 30 & 50  & 0.6864(1) & 0.01068(17) \\
	\hline 
3.4 & $12^4$ & 3000 & 30 & 70  & 0.6980(2) & 0.00662(29)  \\
    & $16^4$ & 2500 & 30 & 60  & 0.6979(1) & 0.00724(13) \\
	\hline
3.5 & $12^4$ & 2500 & 50 & 100 & 0.7086(2) & 0.00364(22)  \\
    & $16^4$ & 2500 &100 & 50  & 0.7086(1) & 0.00450( 8) \\
	\hline
3.6 & $12^4$ & 3000 & 30 & 70  & 0.7182(2) & 0.00164(14)  \\
    & $16^4$ &  $-$ & 26 & $-$ &  $-$      & 0.00271(21) \\
        \hline
\end{tabular}
\end{table}
\begin{table}[htb]
\renewcommand{\arraystretch}{1.3}
	\caption{String tension $\sigma_T$
                  and the corresponding lattice spacing, 
                  as  extracted from
                 linear-plus-Coulomb fits to $V(R,T={\rm fixed})$.} 
	\protect\label{fits}
\begin{tabular}{ccccc}
	\hline
	$\beta$ &$ T $ & $\sigma_{T}$ &$\sigma_{T+1}$ & a (fm) \\
	\hline
	3.1 & 3 & 0.0985( 8) & 0.0965(13)  &  0.141( 6)   \\
	\hline
	3.3 & 4 & 0.0483(11) & 0.0476(16)   &  0.098(11)   \\
	\hline
	3.4 & 5 & 0.0378( 9) & 0.0377( 7) &  0.087(11)   \\
	\hline
	3.5 & 5 & 0.0278( 6) & 0.0277( 7) &  0.075( 8)   \\
	\hline
\end{tabular}
\end{table}
\begin{table}[htb]
\renewcommand{\arraystretch}{1.3}
	\caption{ Runs for string tension determination.
 $N_{\rm iter}$ iterations of APE smearing 
with parameter $c_0$ are used.}
	\protect\label{str_run}
\begin{tabular}{ccccccccc}
	\hline
	$\beta$  
      & lattice  
      & $N_{\rm therm}$
      & $N_{\rm measur}$ 
      & $N_{\rm separ}$ 
      & $c_0$ 
      & $N_{\rm iter}$
      & $ < \Box >$ 
      & $\sigma$ \\    
	\hline
3.1 & $16^4$ & 2500 & 200 & 40  & 8 & 20 & 0.6581(1) & 0.0985(20)\\
	\hline
3.2 & $16^4$ & 2500 & 200 & 40  & 6 & 20 & 0.6732(1) & 0.0702(20) \\
	\hline
3.3 & $16^4$ & 2500 & 200 & 20  & 6 & 20 & 0.6862(1) & 0.0483(19) \\
	\hline
3.4 & $16^4$ & 2500 & 200 & 20  & 5 & 20 & 0.6980(1) & 0.0378( 5)\\
	\hline
3.5 & $16^4$ & 2000 & 200 & 20  & 5 & 20 & 0.7086(1) & 0.0278( 7)\\
	\hline
3.6 & $16^4$ &  $-$   & 200 & 20  & 5 & 20 & 0.7184(1) & 0.0195(10)\\
	\hline
\end{tabular}
\end{table}
\begin{table}[htb]
\renewcommand{\arraystretch}{1.3}
	\caption[dummy]{Asymptotic scaling 
                  for SU(2) string tension with improved action.
                  Different coupling schemes are used to get $a(\beta)\Lambda$
               from \myr(L0): $\a_{0}  = {5\over 3} (\pi\beta)^{-1} $,
                $ \a_{\rm TI}  = \a_{0} \langle\Box\rangle^{-1}$, and 
                $ \a_{s}= -{\log\langle \Box \rangle\over 1.726}$, from
             \myr(app5a). 
  }
	\protect\label{schemes}
\begin{tabular}{ccccccc}
	\hline
	$\beta$  
      & lattice
      & $\sqrt{\sigma}\over\Lambda_{0}$
      & $\sqrt{\sigma}\over\Lambda_{\rm TI}$
      & $\sqrt{\sigma}\over\Lambda_{\rm s}$      
      & $ \a_{s}\over \a_0 $
      & $ \a_{s}\over \a_{\rm TI} $
      \\    
	\hline
 2.7 & $12^4$  &  18.86(52)  &  3.76(10)  &  4.35(12) & 1.732 & 0.935  \\
        \hline
 2.9 & $12^4$  &  18.84(47)  &  3.91(10)  &  4.64(12) & 1.609 & 0.936  \\
        \hline
 3.1 & $16^4$  &  17.75(18)  &  3.82( 4)  &  4.72( 5) & 1.519 & 0.932  \\
	\hline
 3.2 & $16^4$  &  17.38(25)  &  3.79( 5)  &  4.76( 7) & 1.486 & 0.931  \\
	\hline
 3.3 & $16^4$  &  16.73(33)  &  3.68( 7)  &  4.68( 9) & 1.457 & 0.931  \\
	\hline
 3.4 & $16^4$  &  17.17(11)  &  3.81( 3)  &  4.89( 3) & 1.433 & 0.932  \\
	\hline
 3.5 & $16^4$  &  17.10(22)  &  3.81( 5)  &  4.93( 6) & 1.411 & 0.933  \\
	\hline
 3.6 & $16^4$  &  16.63(43)  &  3.72(10)  &  4.85(12) & 1.392 & 0.934  \\
\hline
\end{tabular}
\end{table}

\begin{figure}[htb]
\begin{center}
\mbox{\epsfig{file=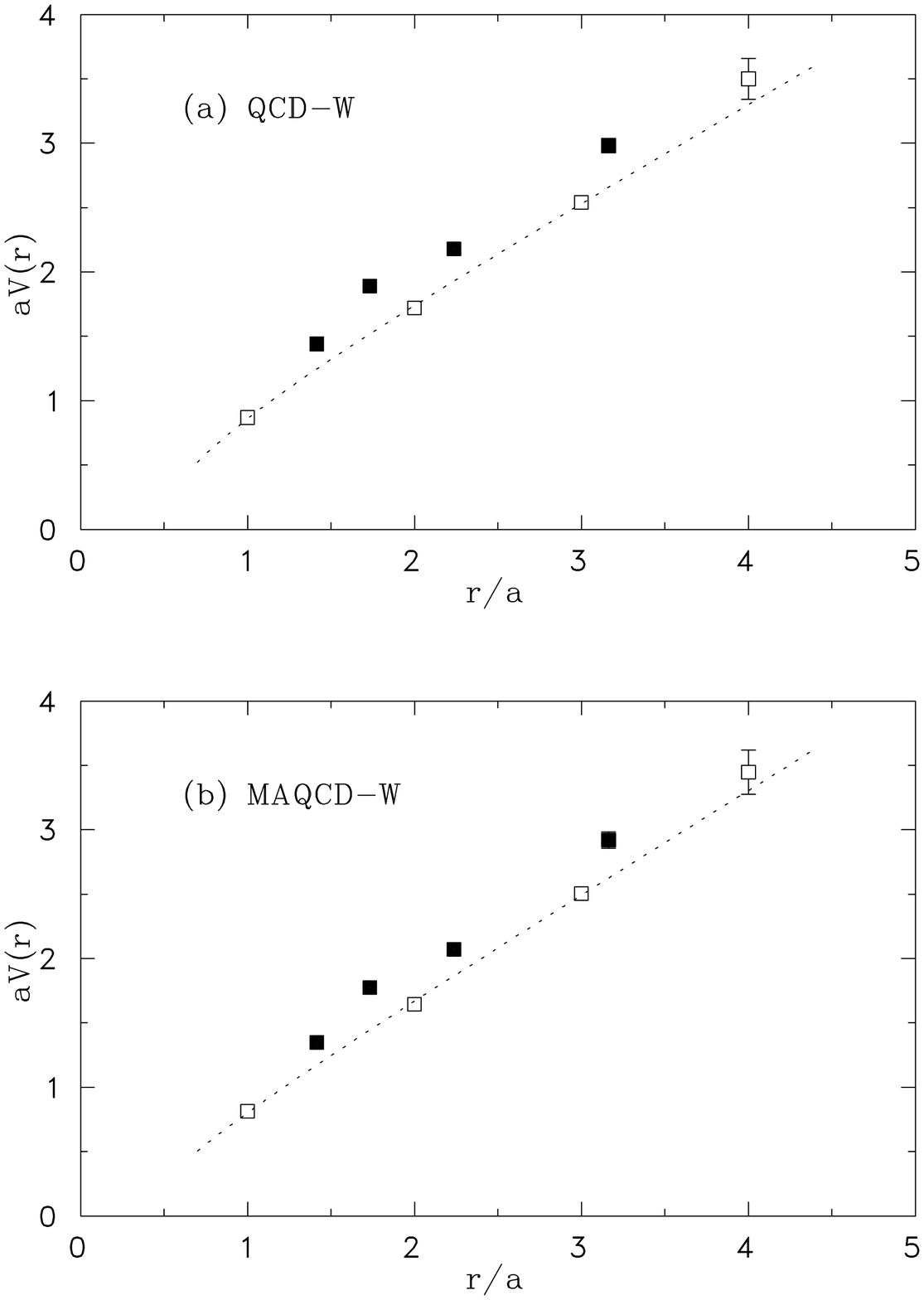,width=.75\textwidth,angle=0}}
\caption[dummy]{ 
         The on-($\Box$) and off-axis (\protect{\rule{1.7mm}{1.7mm}}) 
         nonabelian (a) and abelian (b) potential, from
         $6^4$ lattices at $\beta=1.7$, using the Wilson action.
      }
\label{W6b17}
\end{center}
\end{figure}
\begin{figure}[htb]
\begin{center}
\mbox{\epsfig{file=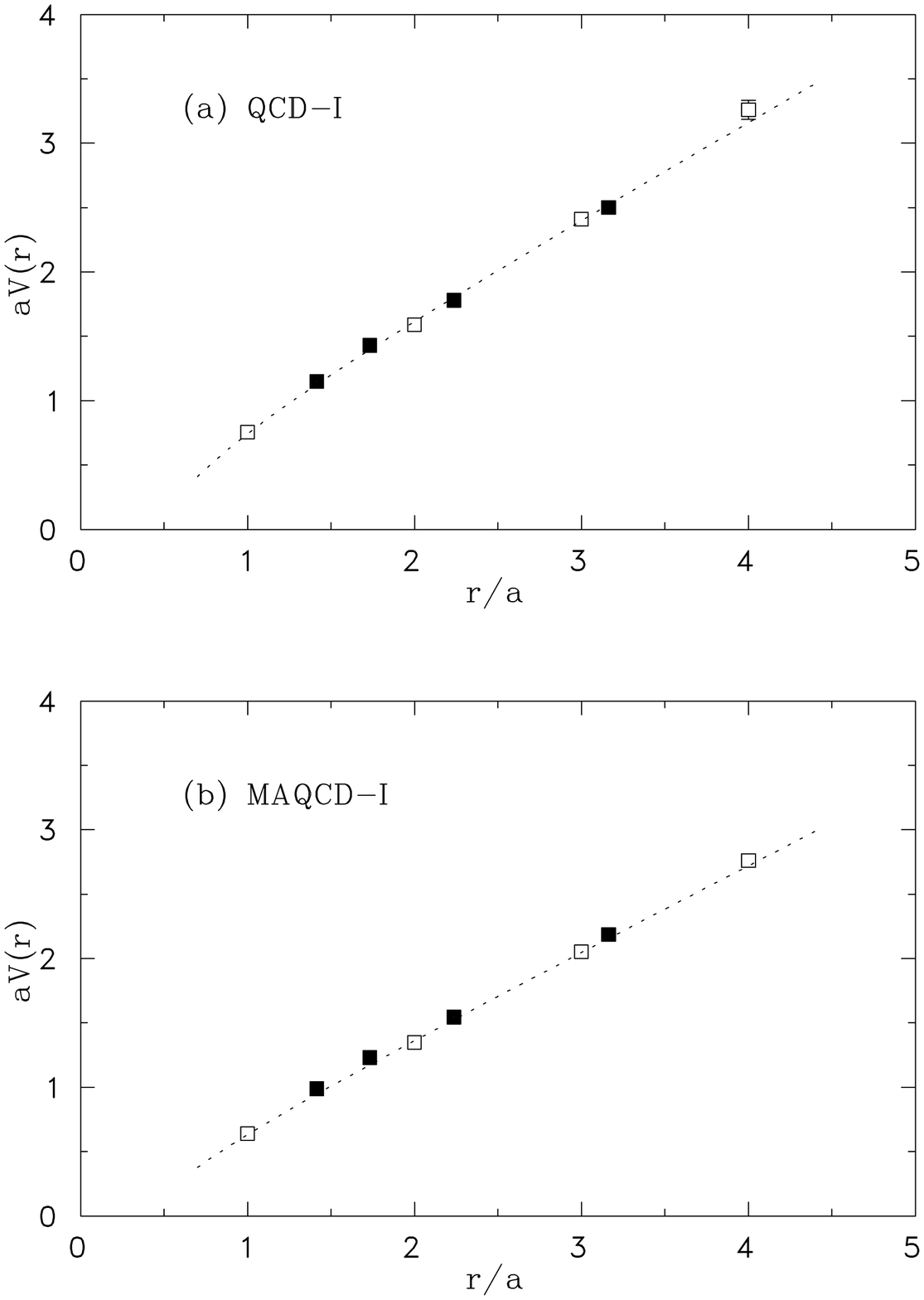,width=.75\textwidth,angle=0}}
\caption[dummy]{ 
               As in Fig.\ \ref{W6b17} but using the
               improved action, at $\beta=2.4$, where the lattice 
               spacing is approximately the same.
              }
\label{I6b24}
\end{center}
\end{figure}
\begin{figure}[htb]
\begin{center}
\mbox{\epsfig{file=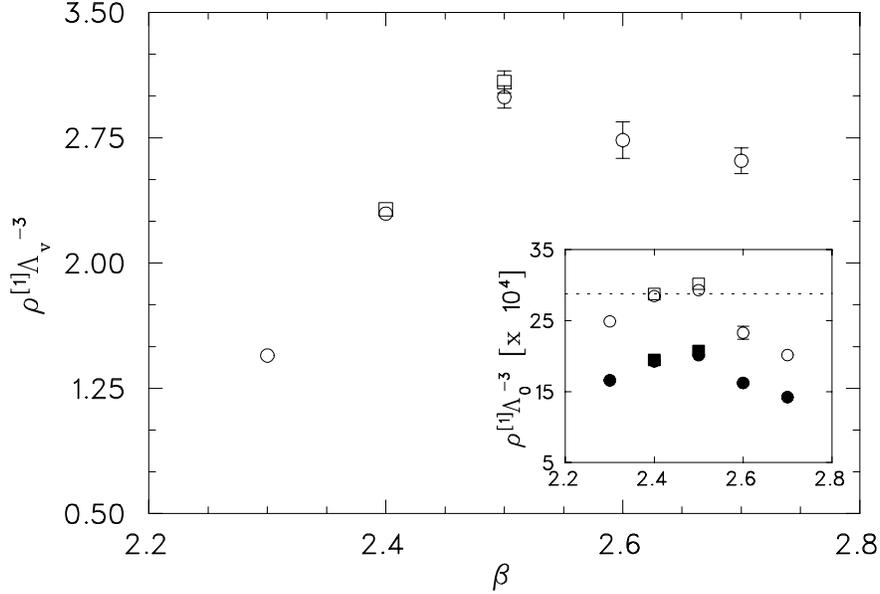,width=.7\textwidth,angle=90}}
\caption[dummy]{Asymptotic scaling test for the density 
                of elementary
              monopoles in MA projection 
         using the Wilson action (MAQCD-W).
             Data from Ref.~\cite{Born4d} (circles) and
               Ref.~\cite{Debbio4d} (squares). 
               Open symbols: 2-loop
           evolution with the $\a_{\rm V}$ coupling
	     [\equ(Wplaq2)] for the main graph, and
          the bare coupling $\alpha_0$ = $(\pi\beta)^{-1}$
          for the inner graph. Solid symbols: 3-loop evolution 
         (bare coupling only). The dashed line 
         corresponds to the scaling curve,
            $\rho=[66\Lambda]^3$, quoted from Ref.~\cite{Born4d}.
       }
\label{Wscaling}
\end{center}
\end{figure}
\begin{figure}[htb]
\begin{center}
\mbox{\epsfig{file=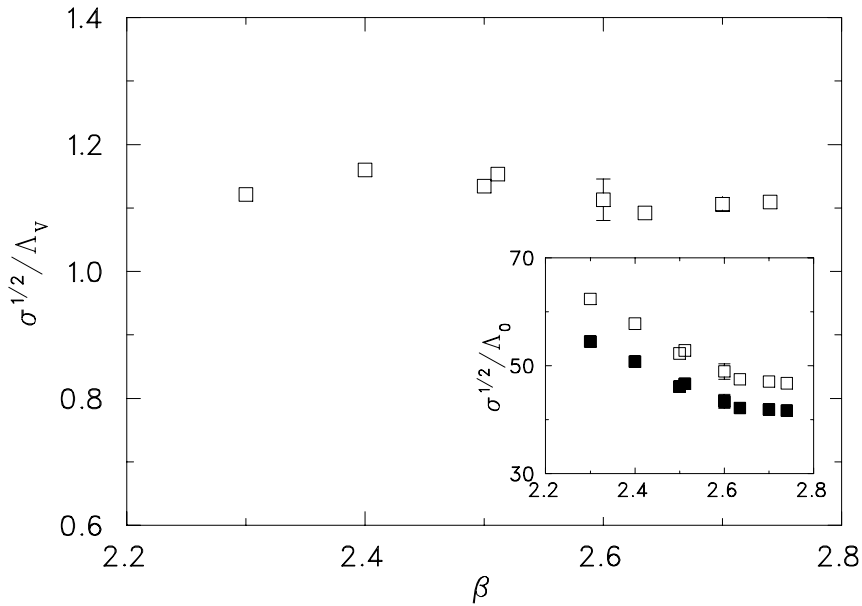,width=.7\textwidth,angle=90}}
\caption[dummy]{Asymptotic scaling test for the
               Wilson action SU(2) string tension. The scale $a\Lambda$ is
                extracted as in Fig.~\ref{Wscaling}. 
              Data from Refs.~\cite{F,Wup,MT}.
       }
\label{s-sc-W}
\end{center}
\end{figure}
\begin{figure}[htb]
\begin{center}
\mbox{\epsfig{file=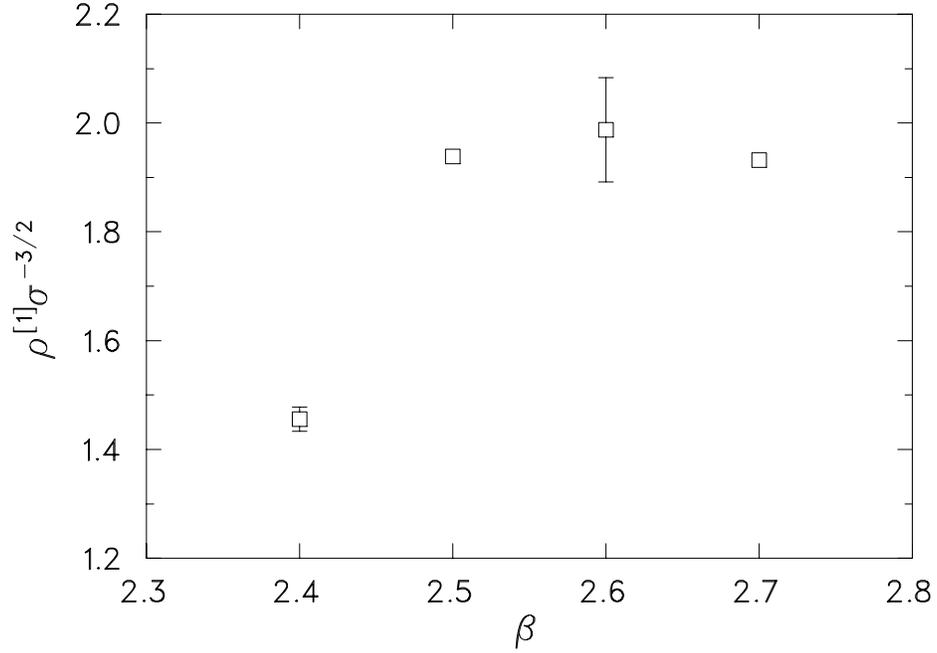,width=.75\textwidth,angle=90}}
\caption[dummy]{Scaling test for the density of elementary
              monopoles against the nonabelian string tension,
             using the Wilson action. 
            Data from Refs.~\cite{F,Born4d,MT}.
       }
\label{ratio-W}
\end{center}
\end{figure}
\begin{figure}[htb]
\begin{center}
\mbox{\epsfig{file=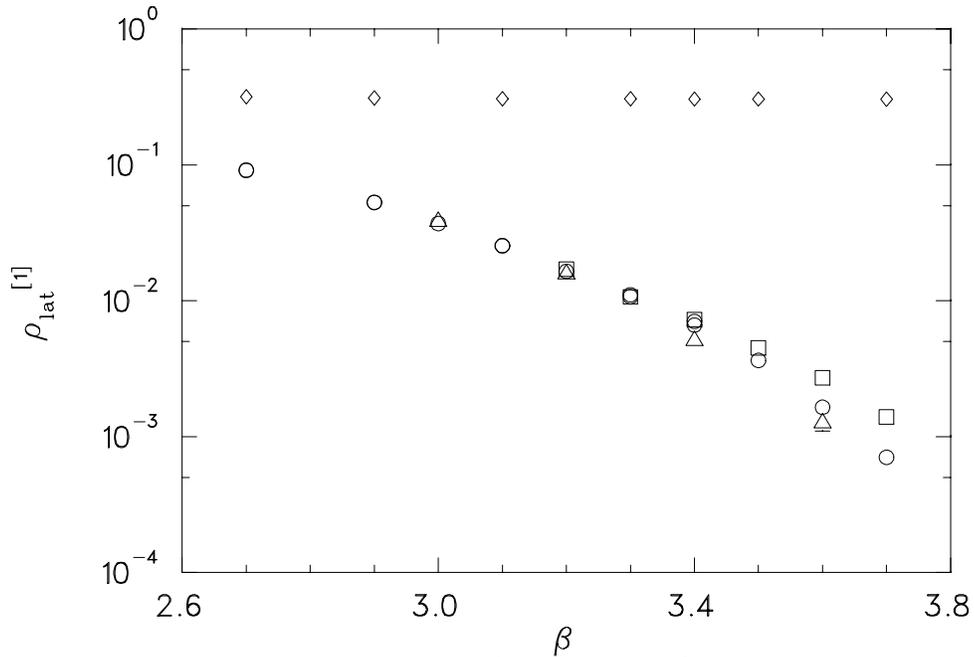,width=.75\textwidth,angle=90}}
\caption[dummy]{ The density of elementary monopoles 
        (in lattice units) in the abelian-projected improved theory
        (APQCD-I). Results shown
         for F12 projection  on $8^4$ lattices 
         ($\Diamond$), and MA projection on
         $8^4$ ($\triangle$),  $12^4$ (o) and  $16^4$ 
         ($\Box$) lattices. 
       }
\label{raw}
\end{center}
\end{figure}
\begin{figure}[htb]
\begin{center}
\mbox{\epsfig{file=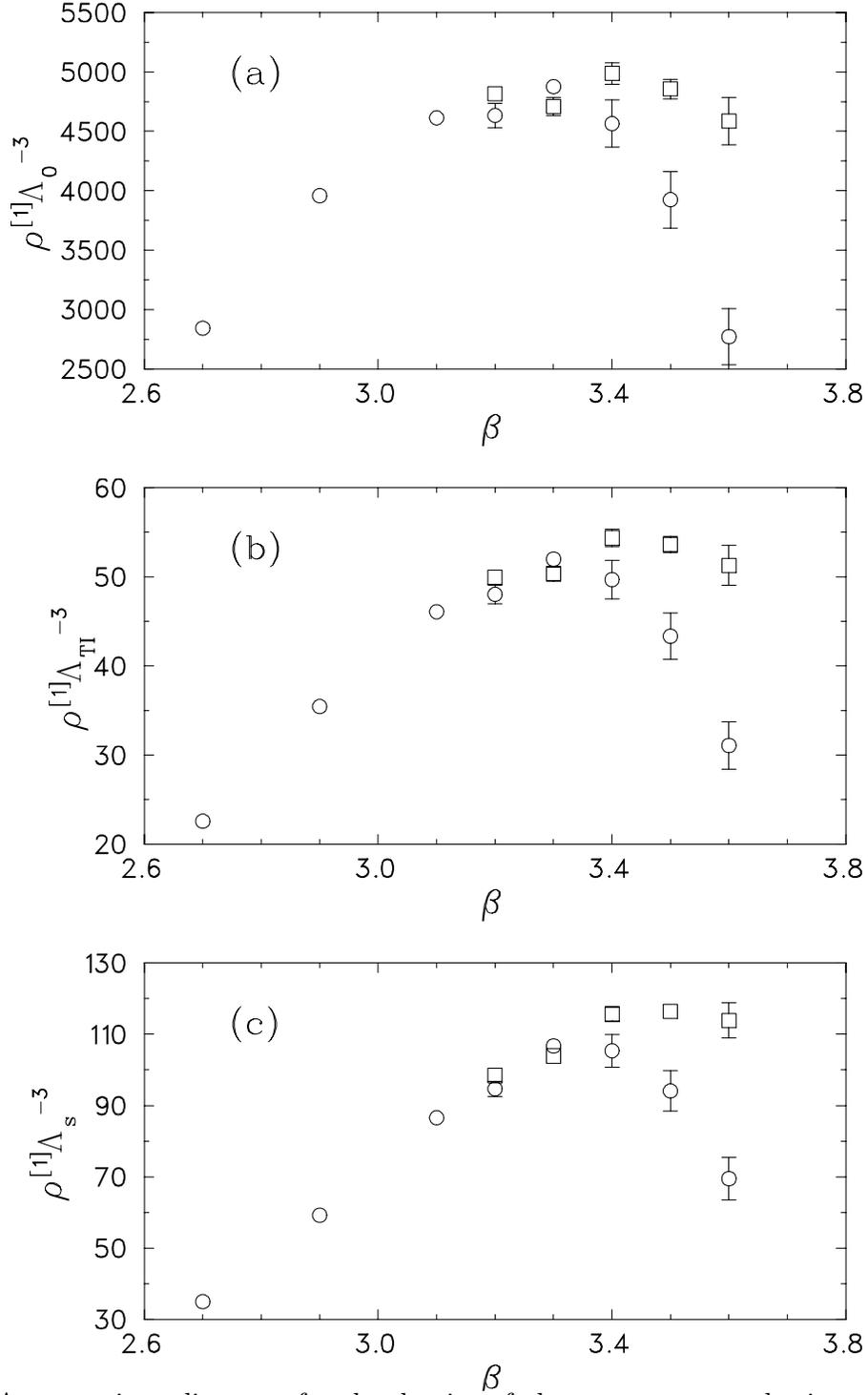,width=.70\textwidth,angle=0}}
\caption[dummy]{ Asymptotic scaling test for the 
          density of elementary monopoles
        in maximal abelian projected SU(2) using the 
         improved action (MAQCD-I).
          The scale $a\Lambda$ is extracted
         (a) from the bare coupling $\a_0$ [\equ(proper)], 
         (b) from the boosted coupling $\a_{\rm TI}=\a_0 u_0^{-4}$,
          and (c) the nonperturbative coupling $\a_{\rm s}$ 
            [\equ(app5a)]. Results shown for
         $12^4$ (o) and $16^4$ ($\Box$) lattices.
       }
\label{mono-scaling}
\end{center}
\end{figure}
\begin{figure}[htb]
\begin{center}
\mbox{\epsfig{file=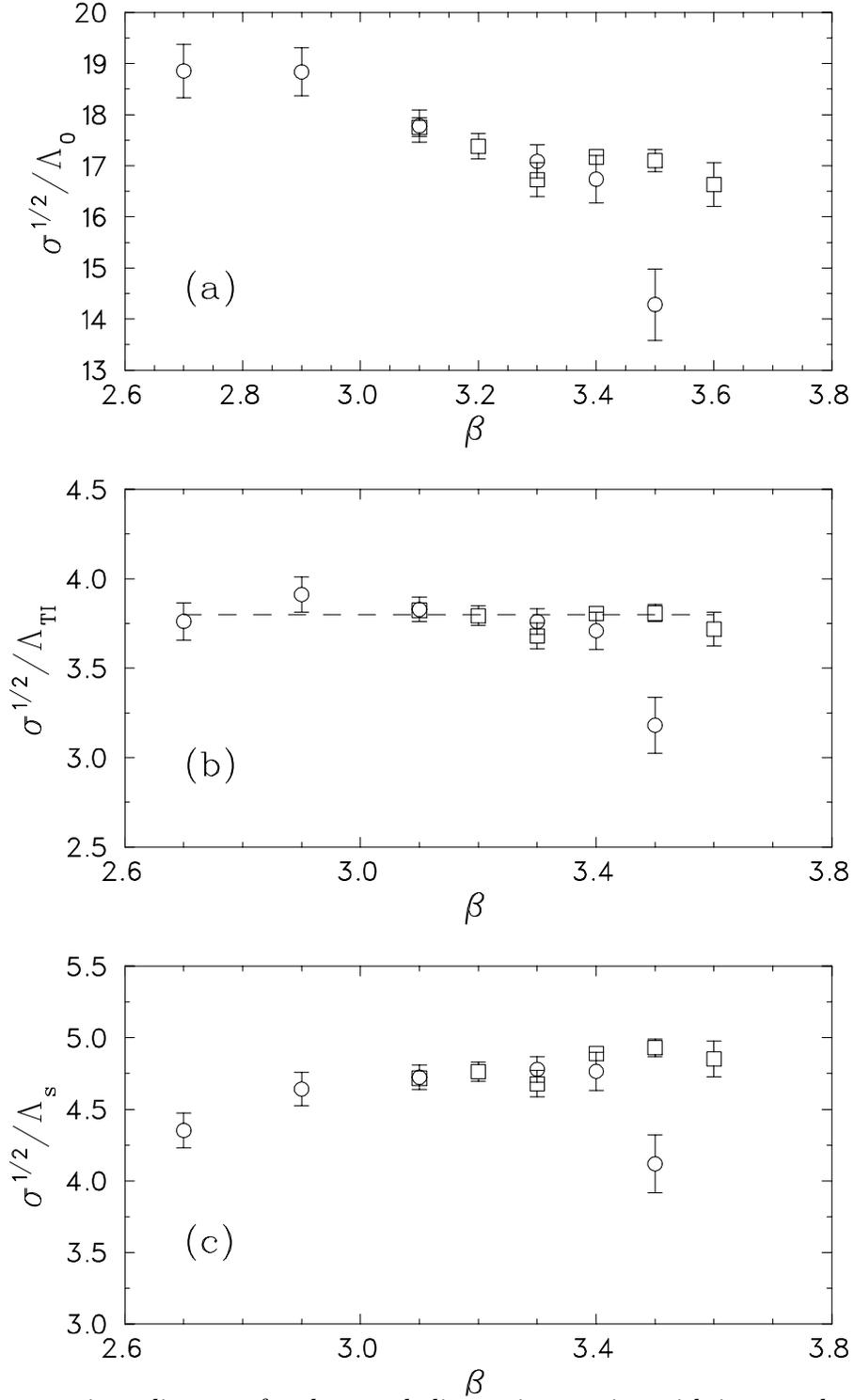,width=.70\textwidth,angle=0}}
\caption[dummy]{Asymptotic scaling test for the nonabelian string
                tension with improved action (QCD-I).
	       The scale is extracted as in Fig.~\ref{mono-scaling}.
                Results are shown for
                 $12^4$ (o) and $16^4$ ($\Box$) lattices.
               The dashed line in (b) corresponds to 
               $3.8\Lambda_{\rm TI}$.
        }
\label{sigma-scaling}
\end{center}
\end{figure}
\begin{figure}[htb]
\begin{center}
\mbox{\epsfig{file=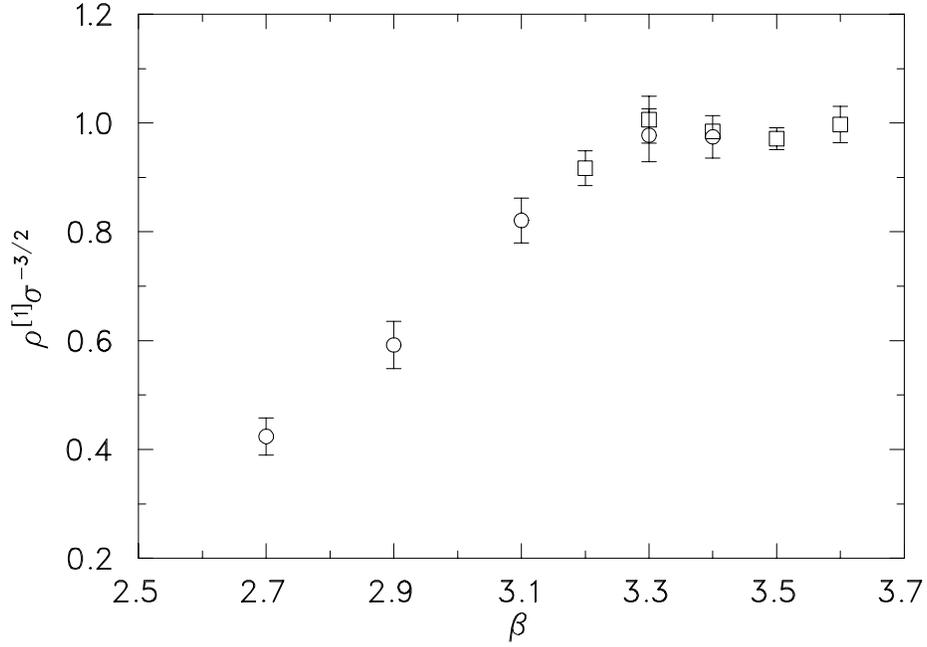,width=.75\textwidth,angle=90}}
\caption[dummy]{Scaling test for the density of elementary
               monopoles against the nonabelian string tension,
               using the tadpole improved action.
               Results are shown for
          $12^4$ (o) and $16^4$ ($\Box$) lattices.
        }
\label{scaling-rs}
\end{center}
\end{figure}
\begin{figure}[htb]
\begin{center}
\mbox{\epsfig{file=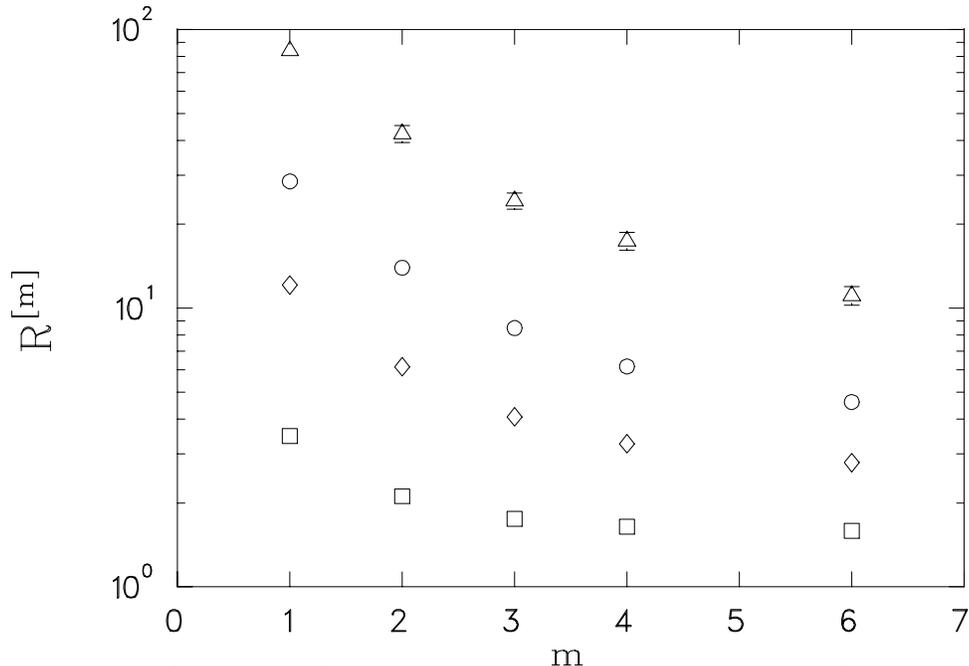,width=.75\textwidth,angle=90}}
\caption[dummy]{ The ratio of densities of extended monopoles 
          between F12 and MA projections using the improved action, 
       as function of the size $m$ of the lattice cube defining the 
          type-II extended monopole. Results from $12^4$ lattices at
          $\beta$ = 2.7 ($\Box$), $\beta$ = 3.1 ($\Diamond$),
          $\beta$ = 3.3 (o), $\beta$ = 3.5 ($\triangle$),
        }
\label{conv}
\end{center}
\end{figure}
\begin{figure}[htb]
\begin{center}
\mbox{\epsfig{file=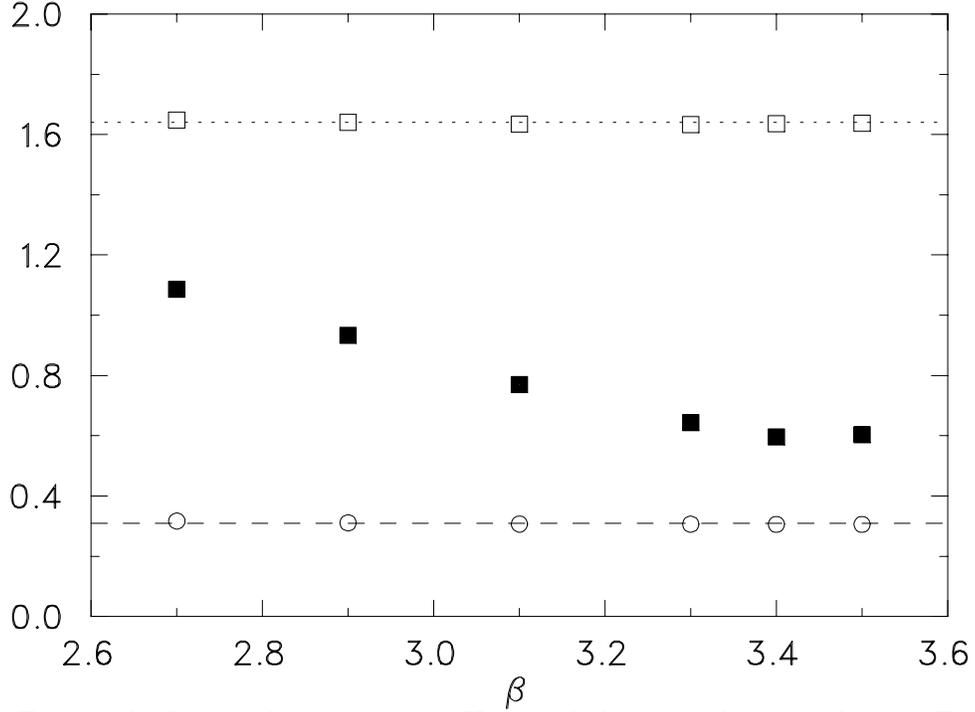,width=.75\textwidth,angle=90}}
\caption[dummy]{Fitting the lattice density of type-II extended
                monopoles according to Eq.~(\ref{f12}):
                $C_{\rm F12}$ (o),
               $\gamma_{\rm F12}$~($\Box$), $\gamma_{\rm MA}$
                (\protect{\rule{1.7mm}{1.7mm}}). Dashed line corresponds
                 to $C$=0.31 and dotted line to $\gamma$=1.64 (see text).
         }
\label{power}
\end{center}
\end{figure}
\begin{figure}[htb]
\begin{center}
\mbox{\epsfig{file=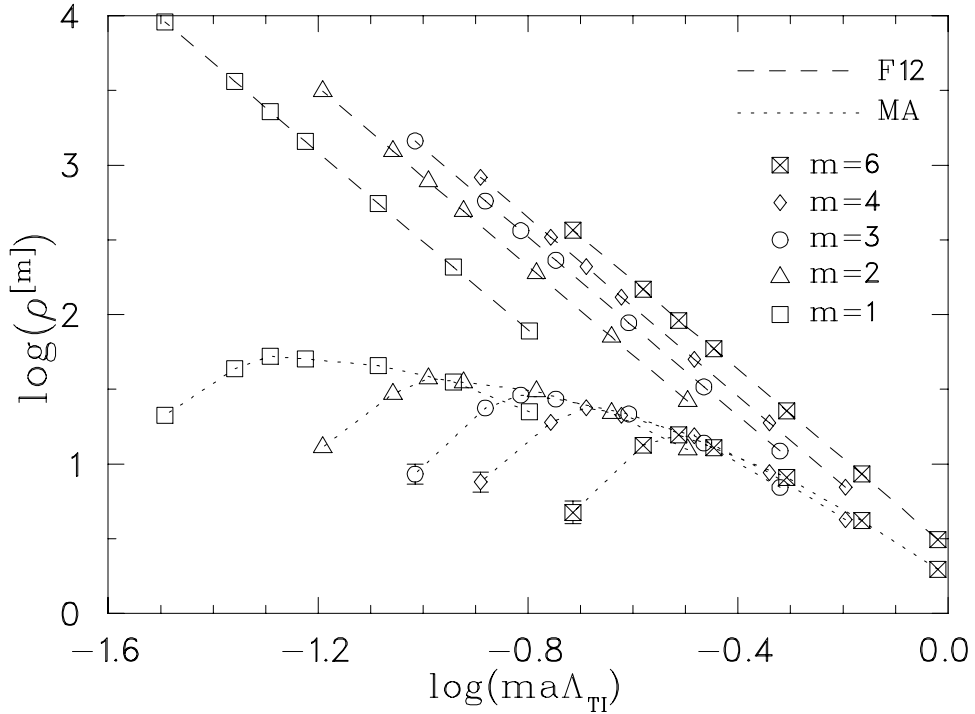,width=.75\textwidth,angle=90}}
\caption[dummy]{The density of extended 
                 monopoles 
          as a function of their size 
           for F12 and MA projections.
       }
\label{univ}
\end{center}
\end{figure}


\begin{references}
\bibitem{beta24} We have used 0.37001(1) for the plaquette
                 at $\beta=2.4$ instead of the value in 
                 Table VII of Ref.~\cite{Wup}.
\bibitem{ImpC} 
an improved Creutz ratio, involving a suitable linear combination
of $W(R+m,R+n)$ Wilson loops with $-2\le m,n\le 1$, 
$|m-n|\le 2$, was proposed in Ref.\ \cite{WW}.
\bibitem{RoyG} For example, one may attempt to improve the
gauge condition [\equ(om)] itself by adding 
a term involving, e.g., length-2 links so as to cancel the 
leading ${\cal O}(a^2)$
error in the discretization of the $D_\mu^0 A^\mu_{\pm}$ = 0
continuum condition (G. Poulis and R. Wensley, in progress).
\end{references}
\end{document}